\documentclass{emulateapj}

\pdfoutput=1 	
\usepackage{epsf,amsfonts,amsmath,amssymb}
\usepackage{float}
\usepackage[pagebackref=false]{hyperref}
\usepackage[all]{hypcap} 
\usepackage{cleveref}
\usepackage{graphicx}
\usepackage{dcolumn}
\usepackage{bm}
\usepackage{multirow}
\usepackage{enumitem} 
\usepackage{xcolor} 
\usepackage[normalem]{ulem} 

\graphicspath{{/home/bnhan/Documents/research/publications/cosmic_twilight_polarimeter_ctp/ctp_implementation_aspects/Figures/}}
\newcommand{\myreferences}{/home/bnhan/Documents/research/publications/cosmic_twilight_polarimeter_ctp/ctp_implementation_aspects/aastex_format/Reference}

\newcommand\T{\rule{0pt}{2.6ex}}   
\newcommand\B{\rule[-1.2ex]{0pt}{0pt}}
\newcommand\tp{\theta,\phi}
\newcommand\tpn{\theta,\phi,\nu}
\newcommand\omeganu{\Omega, \nu}
\newcommand{\diff}{\mathop{}\!\mathrm{d}} 
\newcommand{\tLST}{t_{\rm LST}}

\begin{document}
\setlength{\abovedisplayskip}{5pt}
\setlength{\belowdisplayskip}{5pt}

\title{Assessment of the Projection-induced Polarimetry Technique
  \\ for Constraining the Foreground Spectrum in Global 21 cm Cosmology}

\author{Bang D. Nhan\altaffilmark{1,2,3,7}, David D.
  Bordenave\altaffilmark{1,2,7}, Richard F. Bradley\altaffilmark{1,2,4},
  Jack O. Burns\altaffilmark{3},\\ Keith Tauscher\altaffilmark{3,5}, David
  Rapetti\altaffilmark{3,6}, and Patricia J. Klima\altaffilmark{2}}
 
\altaffiltext{1}{Department of Astronomy, University of Virginia,
  Charlottesville, VA 22903, USA \href{mailto:bnhan@nrao.edu}{bnhan@nrao.edu}}
\altaffiltext{2}{National Radio Astronomy Observatory (NRAO)
  Technology Center (NTC), Charlottesville, VA 22903, USA}
\altaffiltext{3}{Center for Astrophysics and Space
  Astronomy (CASA), Department of Astrophysical and Planetary
  Sciences,\\ University of Colorado, Boulder, CO 80309, USA}
\altaffiltext{4}{Department of Electrical and Computer Engineering,
  University of Virginia, Charlottesville, VA 22903, USA}
\altaffiltext{5}{Department of Physics,
  University of Colorado, Boulder, CO 80309, USA}
\altaffiltext{6}{NASA Ames Research Center,
  Moffett Field, CA 94035, USA}
\altaffiltext{7}{NRAO Grote Reber Doctoral Research Fellow at NTC}

\date{\today}

\begin{abstract}
 Detecting the cosmological sky-averaged (global) 21 cm signal as a
 function of observed frequency will provide a powerful tool to study
 the ionization and thermal history of the intergalactic medium (IGM)
 in the early Universe ($\sim$ 400 million years after the Big
 Bang). The greatest challenge in conventional total-power global 21
 cm experiments is the removal of the foreground synchrotron emission
 ($\sim 10^3$-$10^4$ K) to uncover the weak cosmological signal (tens
 to hundreds of mK), especially since the intrinsic smoothness of the
 foreground spectrum is corrupted by instrumental effects. Although
 the EDGES team has recently reported an absorption profile at 78 MHz
 in the sky-averaged spectrum, it is necessary to confirm this
 detection with an independent approach. The projection effect from
 observing anisotropic foreground source emission with a wide-view
 antenna pointing at the North Celestial Pole (NCP) can induce a net
 polarization, referred as the Projection-Induced Polarization Effect
 (PIPE). Due to Earth's rotation, observation centered at the
 circumpolar region will impose a dynamic sky modulation on the net
 polarization's waveforms which is unique to the foreground
 component. In this study, we review the implementation practicality
 and underlying instrumental effects of this new polarimetry-based
 technique with detailed numerical simulation and a testbed
 instrument, the Cosmic Twilight Polarimeter (CTP). In addition, we
 explore an SVD-based analysis approach for separating the foreground
 and instrumental effects from the background global 21 cm signal
 using the sky-modulated PIPE.
\end{abstract}

\keywords{dark ages, reionization, first stars - techniques:
  polarimetric - methods: observational}

\maketitle

\section{Introduction}
\label{sec:introduction}
Measuring the redshifted 21 cm line corresponding to the hyperfine
transition of neutral hydrogen (HI) has been considered the primary
means to probe the thermal and ionization history of the intergalactic
medium (IGM) in the high-redshift Universe, during the three main
phases known as: the Dark Ages (1,100 $\gtrsim z \gtrsim 30$), Cosmic
Dawn ($30 \gtrsim z \gtrsim 15$), and the Epoch of Reionization (EoR,
$15 \gtrsim z \gtrsim 6$) \citep[e.g, ][]{furlanetto2006cosmology,
  furlanetto2006global, pritchard2010constraining, pritchard201221,
  liu2013global, barkana2016rise}. Development of large
interferometric arrays has been focusing on measuring the power
spectrum of spatial fluctuations in the 21 cm brightness temperature
at the end of EoR \citep[e.g., ][]{parsons2010precision,
  tingay2013murchison, bowman2013science, van2013lofar,
  paciga2013simulation, mellema2013reionization,
  deboer2017hydrogen}. At lower frequencies, observations to constrain
the limits on the Cosmic Dawn using the 21 cm power spectrum have been
attempted by the Low-Frequency Array's Low Band
Antenna~\citep[LOFAR-LBA, ][]{gehlot2018first}.

Meanwhile, for the last decade or so, there have also been efforts to
search for the sky-averaged monopole component of the redshifted 21 cm
signal by using single dipole antennas or compact arrays consisting of
a small number of antenna elements over a large frequency range
(${\sim}50 \le \nu \le 200$ MHz). Some of the current and past
sky-averaged (global) 21 cm experiments include: the Experiment to
Detect the Global EoR Signature \citep[EDGES I \& II,
][]{bowman2008toward, bowman2010lower, monsalve2017calibration}, the
Shaped Antenna Measurement of the Background Radio Spectrum
\citep[SARAS 1 \& 2, ][]{patra2013saras, singh2017first}, the
Broadband Instrument for Global Hydrogen Reionization
Signal~\citep[BIGHORNS, ][]{sokolowski2015bighorns}, the
Large-Aperture Experiment to Detect the Dark Ages \citep[LEDA,
][]{greenhill2015constraining, price2018design}, the Sonda
Cosmol{\'o}gica de las Islas para la Detecci{\'o}n de Hidr{\'o}geno
Neutro \citep[SCI-HI, ][]{voytek2014probing}, and the Probing Radio
Intensity at high-z from Marion \citep[PRIzM, ][]{philip2019probing}.

Theoretical studies have shown that global measurement of the 21 cm
signal can compliment the power spectrum measurements by providing
meaningful physical parameters for the evolution of the IGM
\citep[e.g., ][]{furlanetto2006global, pritchard201221, liu2013global,
  mirocha2013interpreting, mirocha2015interpreting,
  mirocha2017global}. These global experiments are designed to measure
the total power of the sky signal, including both the redshifted 21 cm
background and the foreground synchrotron emission originated from
both Galactic and extragalactic sources.

Among the measurement and systematic errors, the foreground emission
is the strongest contaminant since it is at least four orders of
magnitude ($\sim 10^3$-$10^4$ K) stronger than the cosmological 21 cm
background (tens to hundreds of mK). Nonetheless, the foreground
spectrum is smooth and lacks significant spectral structures
\citep{shaver1999can, tegmark2000foregrounds, petrovic2011systematic,
  kogut2012synchrotron}, whereas the global 21 cm signal is expected
to vary on the scale of tens of MHz. Motivated by its spectral
smoothness, the foreground spectrum is typically approximated by a
polynomial function to separate it from the global 21 cm features. For
example, one common parametrization is the log-log polynomial between
the brightness temperature and frequency
\citep{pritchard2010constraining, bowman2010lower, harker2012mcmc},
\begin{equation}
\log \hat{T}_{fg}(\nu) = \sum_{n=0}^{m>0}c_n\left(\log \nu\right)^n,
\label{eq:log_poly}
\end{equation}
where $\hat{T}_{fg}(\nu)$ is the estimated foreground spectrum with
polynomial coefficient $c_n$ of order $n$ up to a maximum order
$m$.

In principle, the background global 21 cm signal can then be
determined by simultaneously fitting and subtracting the approximated
foreground spectrum from the observed spectrum, after calibrating for
instrumental effects. However, the order of the fitting polynomial can
alter the spectral structures in the residual spectrum. To mitigate
this, using a frequency-independent antenna model,
\cite{rao2017modeling} have shown that approximating the foreground
spectrum with maximally smooth (MS) polynomials can produce a fit more
resilient to the fitting order, and thus preserve the background
signal structure better.

Nonetheless, spectral smoothness in the foreground spectrum is highly
susceptible to corruptions by frequency-dependent antenna beam
patterns (or antenna beam chromaticity) and other instrumental
systematics, thus complicating the foreground fitting. Some studies
\citep{bernardi2015foreground, mozden2016limits,
  monsalve2017calibration} have suggested that beam chromaticity is
not anticipated to compromise the extraction of the background signal
if certain level of beam smoothness can be achieved through special
antenna designs and modeling. In practice, obtaining a true
frequency-independent beam over a large frequency range can be
challenging due to intrinsic characteristics of a broadband
antenna. Any uncorrected residual spectral structures from the beam
can still bias the recovered cosmological signal.

In recent results from EDGES II, \cite{bowman2018absoprtion} have
identified an absorption feature centered at about 78 MHz which has
the potential to be the cosmological signal. However, there is concern
as to whether this feature is a byproduct from fitting errors
\citep{hills2018concerns} or potential instrumental artifacts such as
absorption features induced by a resonant mode excited in the antenna
ground plane \citep{bradley2019ground}. As this active debate in the
community continues, it is imperative to explore a different
measurement approach for potential follow-up observations.
  
In \cite{nhan2017polarimetric} (hereinafter NB17), using simulations
with an idealized frequency-independent circular Gaussian beam, we
demonstrated that the dynamic modulation of foreground emission when
pointing the antenna at a celestial pole can imprint unique signatures
in the temporal waveforms of the net polarization. This polarization,
referred to as the Projection-Induced Polarization Effect (PIPE), is
induced by geometrical projection of the anisotropic distribution of
foreground sources on the antenna plane when the signal is coupled to
the antenna beam. Importantly, since the isotropic cosmological
background \citep[spatial fluctuations with angular scale $<
  2^{\circ}$, ][]{bittner2011measuring} does not produce a net
polarization, NB17 showed that Fourier decomposition of the unique
waveforms in the sky-modulated PIPE at each observed frequency helps
to reconstruct a copy of the foreground spectrum without invoking any
polynomial fit as in the conventional approach.
  
While a variety of implementation aspects are discussed in NB17, the
PIPE simulations there lack realistic instrumental systematics, such
as beam chromaticity, environmental effects on beam patterns, and
observational effects when observing at a lower latitude other than
the geographic poles to achieve foreground modulation on the PIPE. In
this study, we extend that framework to more in-depth numerical
simulations to further evaluate the proposed technique using
semi-realistic antenna beam models acquired from the computational
electromagnetic (CEM) simulation software, CST\footnote{Computer
  Simulation Technology, \texttt{https://www.cst.com}}.
  
In conjunction, a prototype instrument, the Cosmic Twilight
Polarimeter (CTP), is presented as a working testbed for implementing
the network-theory based calibration scheme and observation strategy
for the sky-modulated PIPE. In this study, we also explore the use of
a Singular Value Decomposition (SVD) based algorithm to separate the
foreground signal and complex instrument systematics from the
background 21 cm signal by simultaneously constraining all four PIPE
Stokes parameters.

The rest of the paper is organized as follows. In
\autoref{sec:induced_polarization}, we review the general approach of
the PIPE and its dynamic modulation using Mueller algebra
formulation. Our PIPE simulations, including major instrumental
effects from the antenna beam, are presented
in~\autoref{sec:pipe_sim}. This is followed by instrumentation details
and calibration procedures for the CTP prototype in
\autoref{sec:ctp_experiment}, where preliminary observation is also
analyzed and compared with simulated data. Additionally,
implementation aspects and mitigation strategies are discussed in
\autoref{sec:other_systematics}. Implications of the sky-modulated
PIPE on extracting the global 21 cm signal in the use of SVD, to
account for a more realistic observational setting, are examined in
\autoref{sec:21cm_extraction_implications}. Finally, we conclude this
work in \autoref{sec:conclusion_future_work} by summarizing the
lessons learned and near-term plans for further improvements on the
CTP instrument and evaluating the PIPE approach.

\section{Projection-induced Polarization Effect}
\label{sec:induced_polarization}
\subsection{Mathematical Formalism Revisited}
\label{sec:math_formalism}
By considering the incoming broadband radio signal to be
quasi-monochromatic, the observed induced polarization at each
direction and frequency can be represented by a Stokes vector,
$\bm{S}_{\rm obs}(\tpn)$, consisting of four Stokes parameters,
$(I,\ Q,\ U,\ V)_{(\tpn)}$. As an extension to the Jones matrix used
to characterize the antenna response in NB17, in this study the
Mueller matrix is adopted to describe the PIPE. The $4\times4$ antenna
Mueller matrix, $\bm{M}_{\rm ant}(\tpn)$, describes the coupling of
the Stokes vector of an incoming signal source to the antenna response
to produce the observed Stokes vector as follows,
\begin{equation}
  \bm{S}_{\rm obs}(\tpn) = \bm{M}_{\rm ant}(\tpn)\bm{S}_{\rm
    src}(\tpn),
  \label{eq:stokes_mueller_transform}
\end{equation}
where $\bm{M}_{\rm ant}(\tpn)$ is computed as an outer product of the
antenna Jones matrix $\bm{J}_{\rm ant}(\tpn)$. The conversion between
the two matrices is summarized in \autoref{appdx:jones2mueller}.

Since the incoming signal is assumed to be unpolarized, the source
Stokes vector reduces to $\bm{S}_{\rm src}(\tpn) = (I_{\rm
  src},\ 0,\ 0,\ 0)_{(\tpn)}$. The observed Stokes vector contains
only contribution of the first column in the Mueller matrix. As a
result, the observed data contain polarized components contributed by
the antenna beam pattern as $(I_{\rm src},\ 0,\ 0,\ 0)_{(\tpn)}
\rightarrow (I_{\rm obs},\ Q_{\rm obs},\ U_{\rm obs},\ V_{\rm
  obs})_{(\tpn)}$. In radio interferometry, this is considered as a
form of polarization leakage~\citep{sutinjo2015understanding}.

For a sky-averaged measurement, the resulting net induced polarization
is a vector sum of the Stokes parameters from all directions. Namely,
the observed Stokes parameters are incoming signal spatially weighted
by the four Mueller components. These net Stokes parameters, sometimes
called Stokes antenna temperatures \citep{piepmeier2008stokes},
observed at each local sidereal time (LST) and frequency channel are
parametrized as,
\begin{align}
  I_{\rm net}(\tLST,\nu) &= \frac{\int_{\Omega}M_{11}(\omeganu)I_{\rm
      src}(\tLST,\omeganu)\diff\Omega}{\int_{\Omega}M_{11}(\omeganu)\diff\Omega} \label{eq:global_stokes_I},\\
  Q_{\rm net}(\tLST,\nu) &= \frac{\int_{\Omega}M_{21}(\omeganu)I_{\rm
      src}(\tLST,\omeganu)\diff\Omega}{\int_{\Omega}M_{21}(\omeganu)\diff\Omega} \label{eq:global_stokes_Q},\\
  U_{\rm net}(\tLST,\nu) &= \frac{\int_{\Omega}M_{31}(\omeganu)I_{\rm
      src}(\tLST,\omeganu)\diff\Omega}{\int_{\Omega}M_{31}(\omeganu)\diff\Omega} \label{eq:global_stokes_U},\\
  V_{\rm net}(\tLST,\nu) &= \frac{\int_{\Omega}M_{41}(\omeganu)I_{\rm
      src}(\tLST,\omeganu)\diff\Omega}{\int_{\Omega}M_{41}(\omeganu)\diff\Omega},
  \label{eq:global_stokes_V}
\end{align}
where $M_{i1}(\omeganu)$ are components of the first column in
$\bm{M}_{\rm ant}(\omeganu)$, with $\diff\Omega =
\sin\theta\diff\theta\diff\phi$. It is also worth noting that Equation
\eqref{eq:global_stokes_I} is analogous to the definition of a
beam-weighted antenna temperature, $T_{\rm ant}(\tLST,\nu)$, for sky
brightness temperature $T_{\rm sky}(\tLST,\omeganu)$
\citep[e.g.,][]{kraus1986radio, wilson2009tools},
\begin{equation}  
  T_{\rm ant}(\tLST,\nu) = \frac{
    \int_{\Omega}F(\omeganu)T_{\rm sky}(\tLST,\omeganu)\diff\Omega}{\int_{\Omega}F(\omeganu)\diff\Omega},
  \label{eq:T_ant_def}
\end{equation}
since $M_{11} = F = (F_X + F_Y)/2$ as shown in Equation
\eqref{eq:appdx_mueller_matrix_expand}, where $F_{X,Y}$ are the
antenna beam patterns for $X$ and $Y$ polarizations.

\subsection{Foreground Modulation on the Induced Polarization}
\label{sec:foreground_modulation}
\begin{figure}
\centering
\includegraphics[width=0.75\columnwidth]{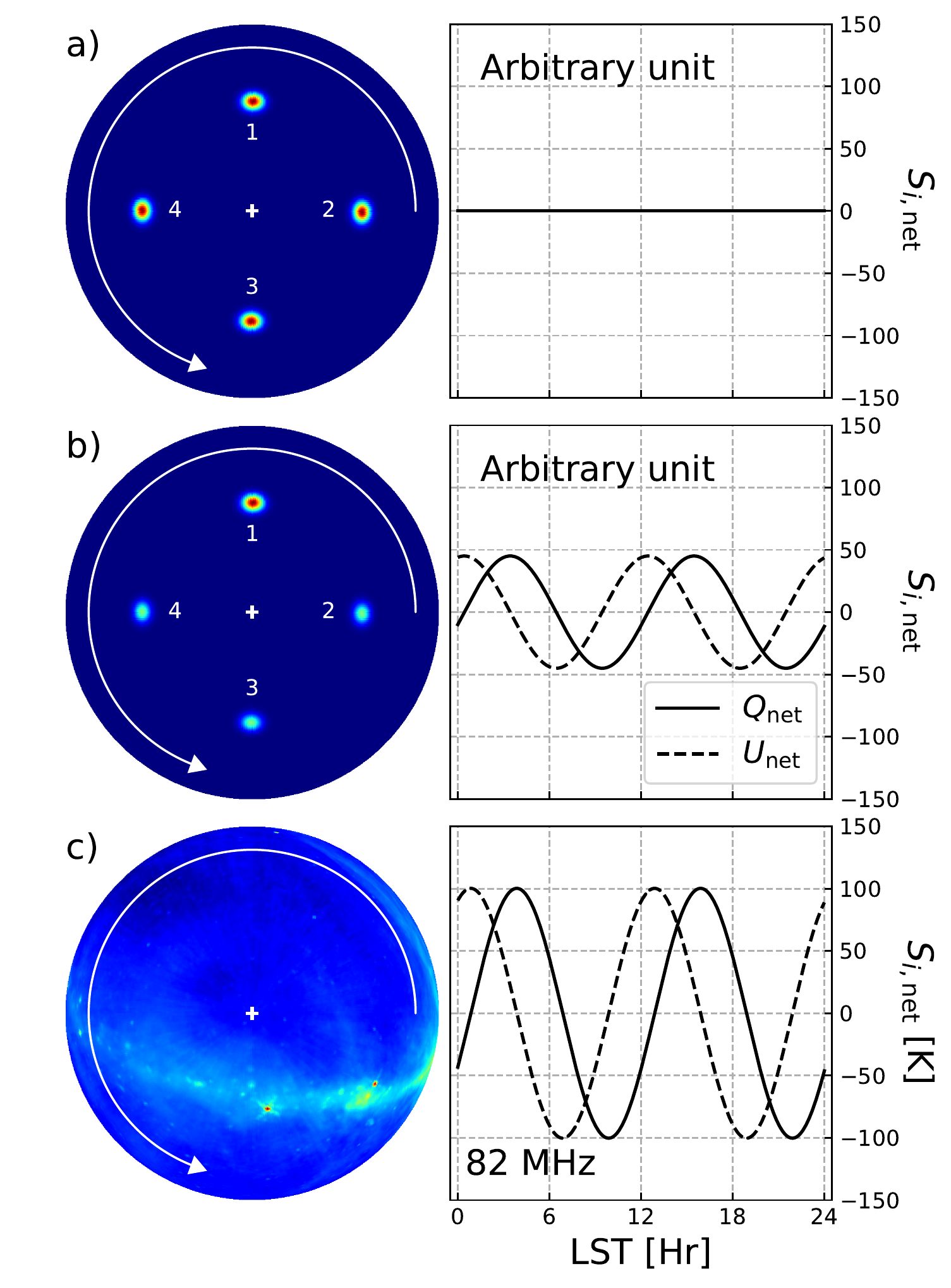}
\caption{Illustrations of three simple scenarios for the PIPE in
  orthographic projection: (a) Four identical point sources at equal
  distance from the center (left) produces zero net polarization
  (right). (b) In a similar situation with four point sources at equal
  distance, a net polarization is produced when one of the point
  sources is stronger than the remaining three. (c) As an example at
  60 MHz, the anisotropic distribution of foreground emission from the
  Haslam map centered at the NCP produces a net polarization which can
  be used to track the foreground, with the signature twice-diurnal
  periodicity.}\label{fig:projection_induced_polarization_illustration}
\end{figure}

For illustration purposes only, the basic rationale of utilizing PIPE
modulation as a means to separate the foreground from the background
signal is presented with three simple scenarios (left column of
\autoref{fig:projection_induced_polarization_illustration}):
  \begin{enumerate}[label=(\alph*)]
  \item An artificial foreground with four point sources with
    identical brightness placed at equal distance from the center of
    the antenna's field of view (FOV),
  \item Similar to (a), but with one of the point sources (Source
    \#1) brighter than the other three,
  \item The Haslam full-sky survey map as the sky brightness
    temperature, scaled from 408 MHz to 60 MHz using a power-law
    function with a constant spectral index $\beta = 2.47$, i.e.,
    $T_{\rm sky}(\nu) = T_{\rm Haslam}(\nu/408\ {\rm MHz})^{-\beta}$,
    where $T_{\rm Haslam}$ is the Haslam map at 408 MHz.
  \end{enumerate}

\begin{figure}
\centering 
\includegraphics[width=0.75\columnwidth]{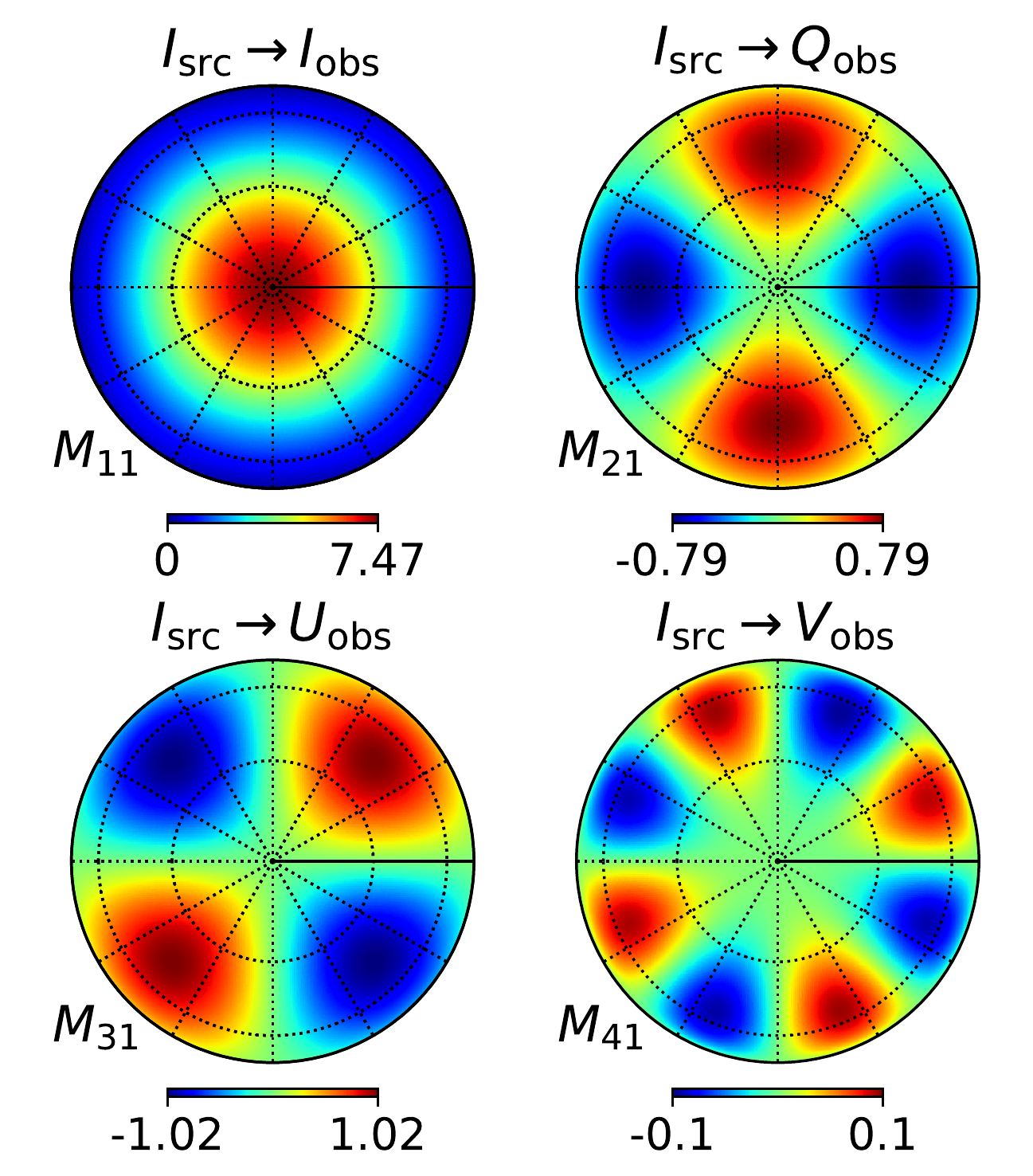}
\caption{Orthographic projection of the first column of the antenna
  Mueller matrix, computed from the Jones matrix obtained using
  magnitudes of simulated farfield gain components of a pair of
  crossed dipoles. The $M_{11}$ component represents the beam pattern
  which produces the total intensity of the observed sky region. The
  $M_{21}$ and $M_{31}$ consist of patterns with a two-fold symmetry
  which respectively contribute to the linear polarizations shown at
  $(0^{\circ}, 90^{\circ})$ and $(+45^{\circ},
  -45^{\circ})$. Meanwhile, the $M_{41}$, with a four-fold symmetry,
  imposes the left-hand and right-hand circular polarizations onto the
  measurement but at a much lower level relative to the other terms,
  as indicated in the color bars in linear units. The grid lines are
  spaced $30^{\circ}$ apart in each direction.}
\label{fig:mueller_2x2_col1_plot_ctp1_skirt0_soil0_grid}
\end{figure}

The complex antenna beams for a pair of simple crossed dipoles above a
finite ground plane are obtained from the CST software, to construct
$\bm{J}_{\rm ant}$. Subsequently, the first column of the
corresponding antenna Mueller matrix is computed using Equation
\eqref{eq:appdx_mueller_matrix_expand} to produce
\autoref{fig:mueller_2x2_col1_plot_ctp1_skirt0_soil0_grid}. The PIPE
for the three scenarios are simulated by computing the beam-weighted
mean of the net Stokes parameters for each corresponding foreground
maps with Equations
\eqref{eq:global_stokes_I}-\eqref{eq:global_stokes_V}.

To dynamically modulate the net polarization, each of the foreground
maps is centered at the antenna's boresight such that a constant FOV
is maintained as the foreground map revolves over 24 sidereal
hours. For the first two artificial point-source maps, the FOV is
aligned at the circumcenter of the four point sources. For the Haslam
map, the antenna pointing is aligned at the North Celestial
Pole\footnote{Similar effect can be achieved for observing at the
  South Celestial Pole (SCP) in the Southern Hemisphere.}  (NCP),
around which the sky revolves.

In scenario (a), due to symmetry in the point source distribution, the
equal but opposite two-fold patterns in $M_{21}$ and $M_{31}$
(\autoref{fig:mueller_2x2_col1_plot_ctp1_skirt0_soil0_grid}) produce a
zero net polarization
(\autoref{fig:projection_induced_polarization_illustration}a,
right). In the second case, since source \#1 is brighter than the
others, the source distribution is uneven. Hence, as the sources
revolve about the $M_{21}$ and $M_{31}$ beams, the amplitude of the
$Q_{\rm net}(t_{\rm LST})$ and $U_{\rm net}(t_{\rm LST})$ are
modulated by a waveform with an angular frequency of twice the
foreground revolution rate
(\autoref{fig:projection_induced_polarization_illustration}b, right).

Similarly, in the last scenario, as asymmetry in the projection of the
Haslam map at 60 MHz revolves about the NCP, the amplitude of the
resulting net polarization is also modulated by the sky rotation to
produce a waveform with a twice-diurnal (two cycles per sidereal day)
periodicity, i.e., $\omega_{Q,U} = 2\omega_{\rm sky}$ where
$\omega_{\rm sky}$ is the sky revolution rate. Also, the phase
difference between $Q_{\rm net}(t_{\rm LST})$ and $U_{\rm net}(t_{\rm
  LST})$ is $\pi/2$ due to the phase shift between $M_{21}$ and
$M_{31}$.

\subsection{Foreground Removal with the Idealized PIPE}
\label{sec:foreground_removal_idealized}
Since the 21 cm background signal has brightness fluctuations on
relatively small angular scales compared to the resolvable foreground
anisotropy, the 21 cm background is equivalently isotropic in a
sky-averaged measurement. Previous sections have illustrated how the
PIPE arises solely from projection of the anisotropic foreground
source distribution on the antenna plane. By pointing the antenna at
the NCP, the modulated waveforms observed in the $Q_{\rm net}(t_{\rm
  LST},\nu)$ and $U_{\rm net}(t_{\rm LST},\nu)$ in principle provide
direct means to constrain the foreground component without the
confusion of any underlying 21 cm background signal mixed in, unlike
the total power spectrum $I_{\rm net}(t_{\rm LST},\nu)$ which contains
both signals.

From the idealized assumptions in NB17, with a spatially constant
foreground of spectral index $\beta$ coupled to a spectrally flat and
symmetric Gaussian beam, the simulated PIPE shows that the foreground
spectrum $T_{\rm fg}(\nu)$ can be reconstructed empirically by Fourier
decomposing waveforms of the Stokes $Q$ or $U$ at each observed
frequency and then compiling magnitudes of the corresponding harmonic
modes. The magnitudes of the power spectral density (PSD) for two
Stokes waveforms are computed at each harmonic mode $n$ to construct
the Stokes spectra as \citep{heinzel2002spectrum},
\begin{equation}
  S_{S_i,n}^{\nu} = \frac{(\Delta t)^2}{s_1^2}
  \left|\sum_{t=1}^{M}w_{\rm BH4}(t)S_i(t)e^{-i2\pi t/(n M \Delta t)}\right|^2,
    \label{eq:stoke_psd}
\end{equation}
where $S_i$ is one of the four Stokes parameters, $w_{\rm BH4}(t)$ is
the four-term Blackman-Harris window function to prevent spectral
leakage, $s_1 = \sum_{t=0}^{M}w_{\rm BH4}(t)$ is the normalization for
discrete data length of $M$ with an averaging interval $\Delta t$.

Hence, in the ideal case, the second harmonic $S_{Q,2}^{\nu}$ (or
$S_{U,2}^{\nu}$) corresponding to the two-fold symmetry in $M_{21}$
(or $M_{31}$) is a scaled version of the foreground spectrum $T_{\rm
  fg}(\nu)$. The input global 21 cm background model $\delta T_{\rm b,
  21cm}(\nu)$ can then be extracted by iteratively scaling and
subtracting the second-harmonic Stokes spectrum from the total sky
spectrum $S_{I,0}^{\nu}$, i.e., $\delta T_{\rm b,21cm}(\nu) = T_{\rm
  sky}(\nu) - T_{\rm fg}(\nu) = S_{I,0}^{\nu} - AS_{Q,2}^{\nu}$, where
$A$ is some best-fitted scaling constant. In
\autoref{sec:21cm_extraction_implications}, we show that a more
sophisticated SVD-based algorithm is needed to constrain the
foreground component in the presence of realistic systematics.

\section{PIPE Simulation with Semi-realistic Beam Models}
\label{sec:pipe_sim}

\subsection{Model Description}
\label{sec:ctp_modeling}
\begin{figure}[!htb]
  \centering
  \includegraphics[width=0.75\columnwidth]{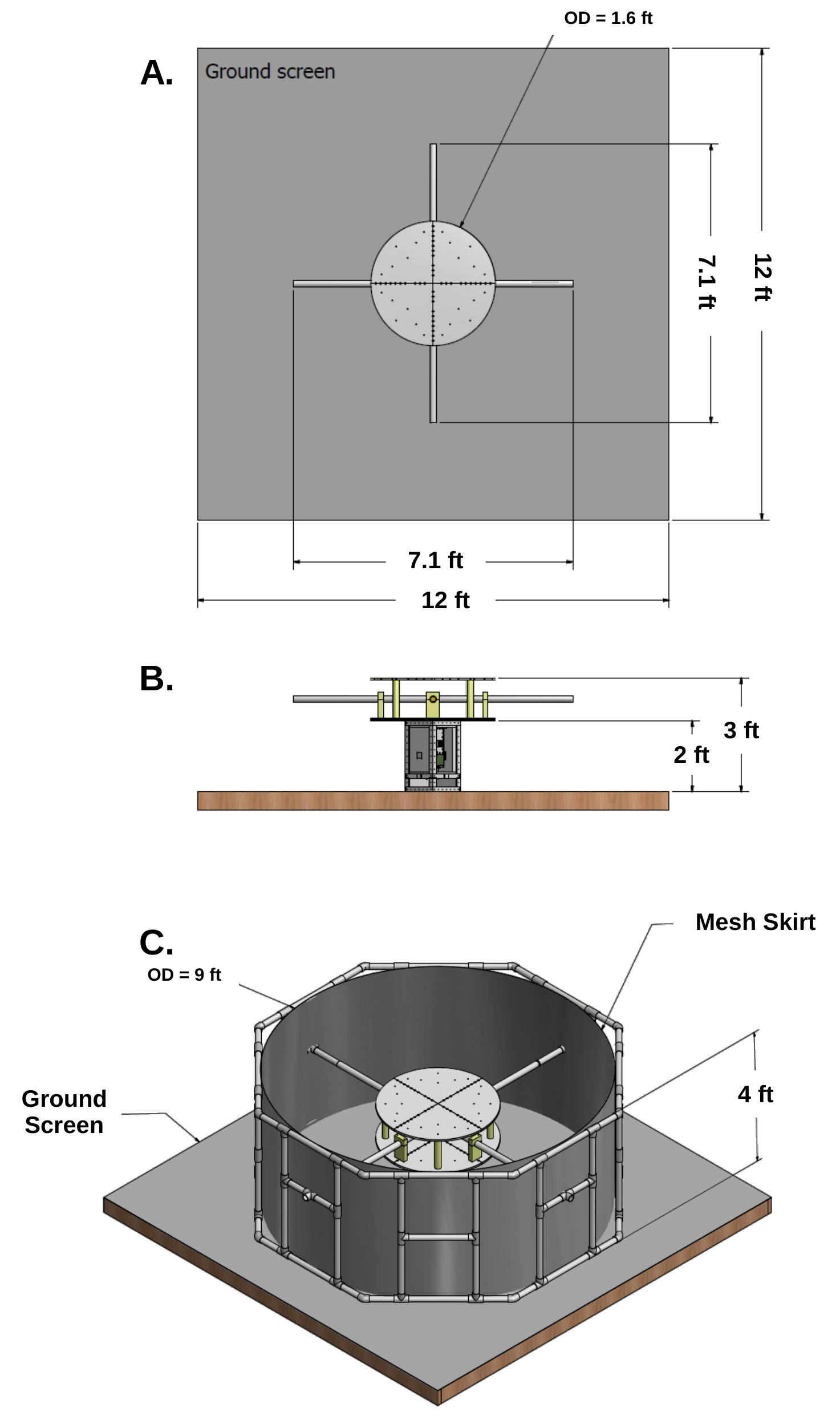}
  \caption{3D model rendering of the sleeved dipole antenna at
    different views. The conductive cylindrical mesh skirt is for
    enhancing the beam symmetry between $E$- and $H$-planes.
  }\label{fig:sleeved_dipole_assembly_ctp}
\end{figure}

Similar to other global 21 cm experiments, the PIPE also requires a
broadband and spectrally smooth antenna. Since preliminary analyses
suggested that a symmetric beam pattern can improve the sensitivity of
sky-modulated PIPE component, a sleeved dipole antenna design with a
concentric conductive skirt was adopted for the study. The antenna
consists of a pair of orthogonal dual-polarized antenna elements
centered between two circular metal plates (or sleeves). The
semi-realistic antenna beams used to compute the Mueller matrices
across the band were obtained from CST simulation computed with
detailed antenna structure model
(\autoref{fig:sleeved_dipole_assembly_ctp}), including a thick ground
soil slab beneath the antenna, as well as antenna's ground plane
tilting orientation to achieve the foreground modulation effect.

Following a similar framework to that described in the previous
section, beam-weighted Stokes parameters for the PIPE were computed
using Equations \eqref{eq:global_stokes_I}-\eqref{eq:global_stokes_V}
for the Haslam sky map scaled to the a band of 60-90 MHz using a
power-law function with a mean\footnote{The foreground spectral index
  is known to be distributed between around 2.4 and 2.6 depending on
  the region in the sky. The effect of a variable $\beta(\tp)$ on the
  PIPE is discussed in \autoref{sec:fg_spectral_index_variations}.}
spectral index of $\beta = 2.47$.

Additionally, the PIPE simulations in this section also account for
obstruction of the Earth's horizon on the FOV of the northern sky
centered at the NCP when observing at lower latitudes. Impacts of the
ground soil and horizon obstruction on the induced polarization,
presented in \autoref{sec:ground_distortions} and
\autoref{sec:horizon_obstruction} respectively, are contrasted to a
fiducial model defined in \autoref{sec:fiducial_model}. To decouple
these effects from other systematics, the simulations assume an
optimal calibration for the electronics, along with the absence of
other environmental variables like RFI and ionospheric distortions.

\subsection{Fiducial Model}
\label{sec:fiducial_model}
The fiducial CST model consists of setting the ground screen of the
sleeved dipole parallel to a finite ground soil slab beneath the
antenna's ground screen. The fiducial model also assumes that the
antenna is located at the Geographic North Pole (GNP) to observe the
NCP at the zenith to avoid the Earth's horizon obstruction.

\begin{figure}
\centering
\includegraphics[width=0.95\columnwidth]{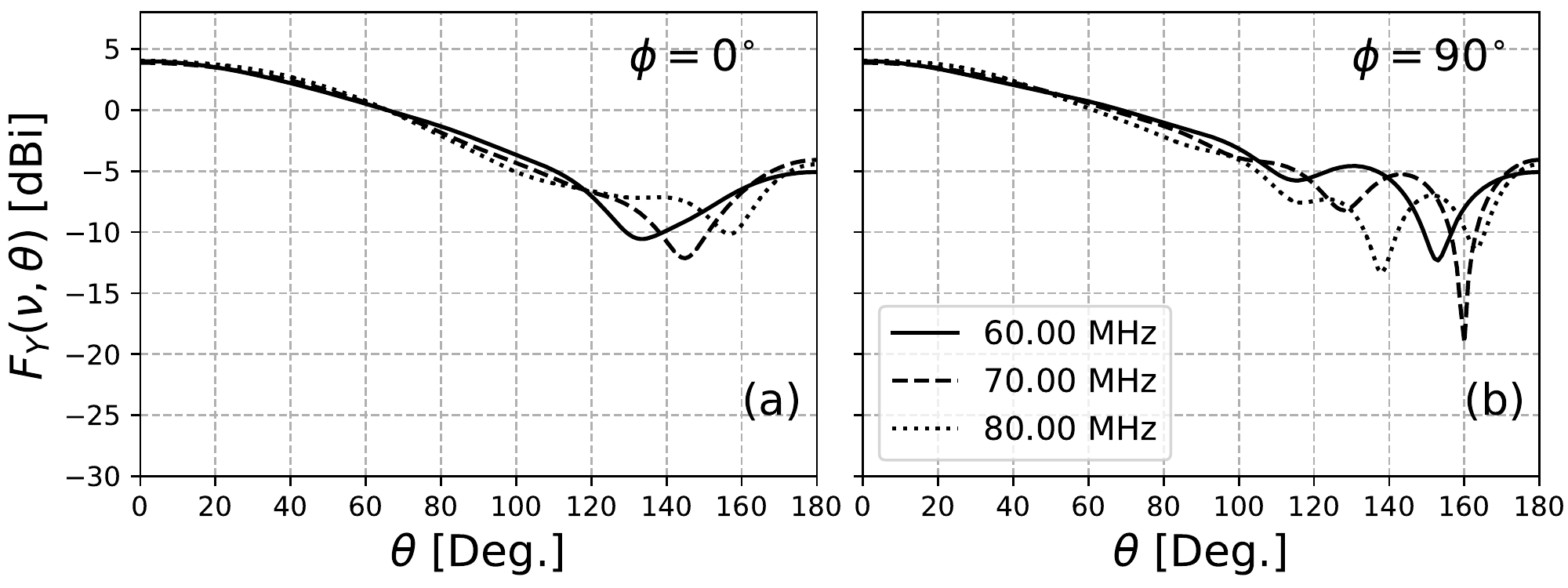}
\caption{Angular plots for the CST beam of the sleeved dipole when the
  antenna is set parallel to the ground. Hence the $E$-plane
  ($\phi=0^{\circ}$) and the $H$-plane ($\phi=90^{\circ}$) of the
  beams are smooth and symmetric. The chromaticity is apparent as the
  beam patterns vary among different frequencies as shown by 60, 70,
  and 80 MHz.}\label{fig:cst_ctp_v1_ideal_polX_slices_log_composite}
\end{figure}

\begin{figure}
\centering 
\includegraphics[width=0.95\columnwidth]{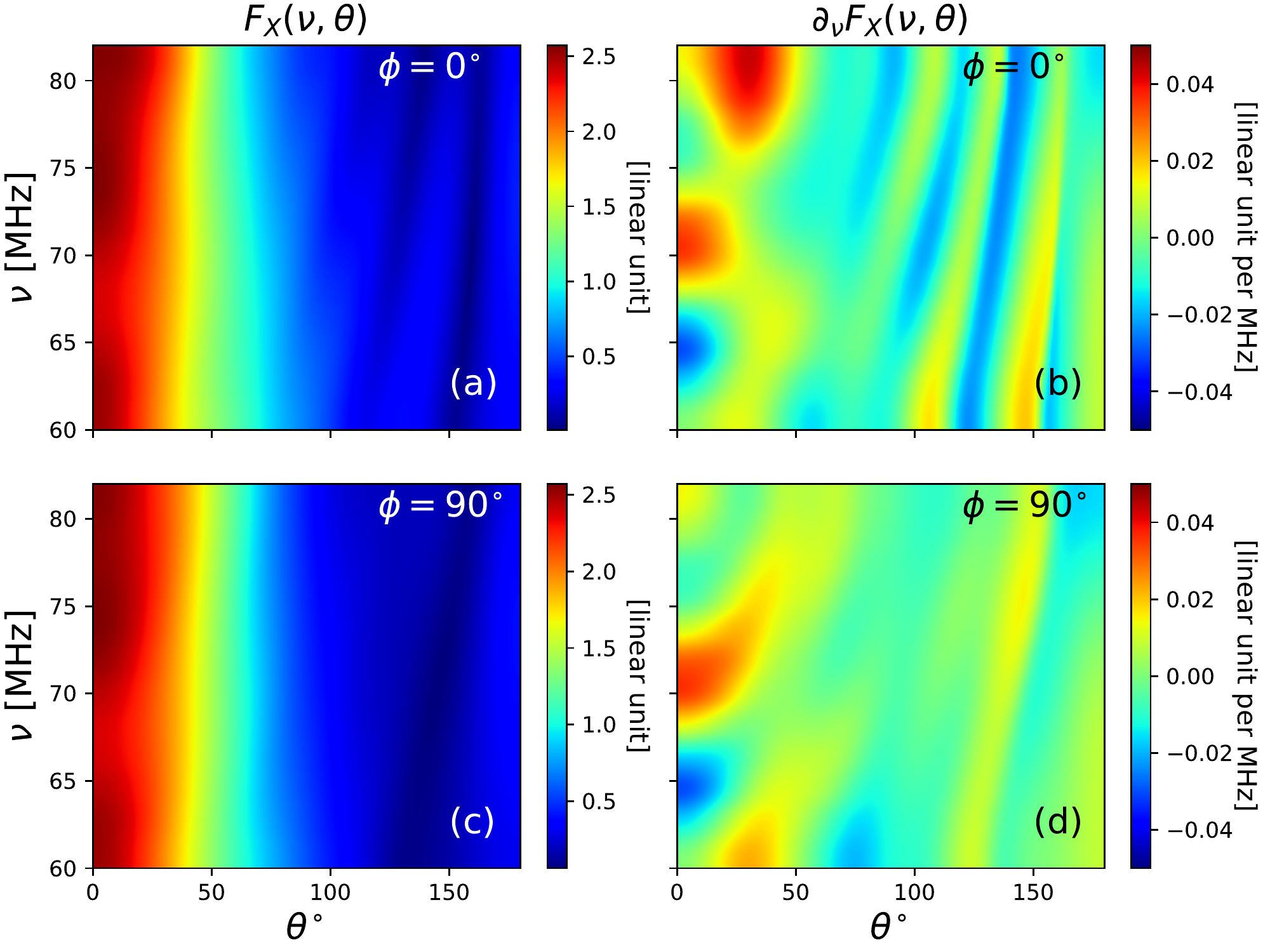}
\caption{(Left) 2D plots of the $E$- and $H$-planes for the
  linear directivity in the $X$-polarization of the fiducial CST beam
  model, $F_X(\tpn)$, which has the finite ground screen parallel to
  the ground soil. (Right) 2D plots of the frequency gradient,
  $\partial_{\nu}F_X(\tpn)$, of the beams on the left panels. The
  strong fringing structures due to interactions between the beam and
  ground are more apparent in the gradient plots, with
  $\left|\partial_{\nu}F(\tpn)\right| \le 0.05$ (linear directivity
  unit per MHz). By symmetry, the beam for $Y$-polarization,
  $F_Y(\tpn)$, shares similar patterns and frequency structures.}
\label{fig:cst_ctp_v1_ideal_polX_gradient}
\end{figure}

The resulting beams $F(\tpn)$ are relatively smooth spatially, with
apparent frequency dependence, as shown in the $E$-plane $(\phi =
0^{\circ})$ and $H$-plane $(\phi = 90^{\circ})$ of polarization $X$ in
\autoref{fig:cst_ctp_v1_ideal_polX_slices_log_composite}.
To further quantify the beam chromaticity of the fiducial beam, we
computed the spectral gradients of the beam patterns
$\partial_{\nu}F(\tpn)$ evaluated at each fixed $(\tp)$. As shown in
\autoref{fig:cst_ctp_v1_ideal_polX_gradient}, the spectral gradients
of the $E$- and $H$-planes plotted side-by-side to their respective
beams are showing large gradient variations with
$\left|\partial_{\nu}F(\tpn)\right| \le 0.05$ (linear directivity unit per
MHz). The significance of these variations is elaborated in
\autoref{sec:ant_beam_chromaticity}.

\begin{figure}[!htb]
  \centering
  \includegraphics[width=0.9\columnwidth]{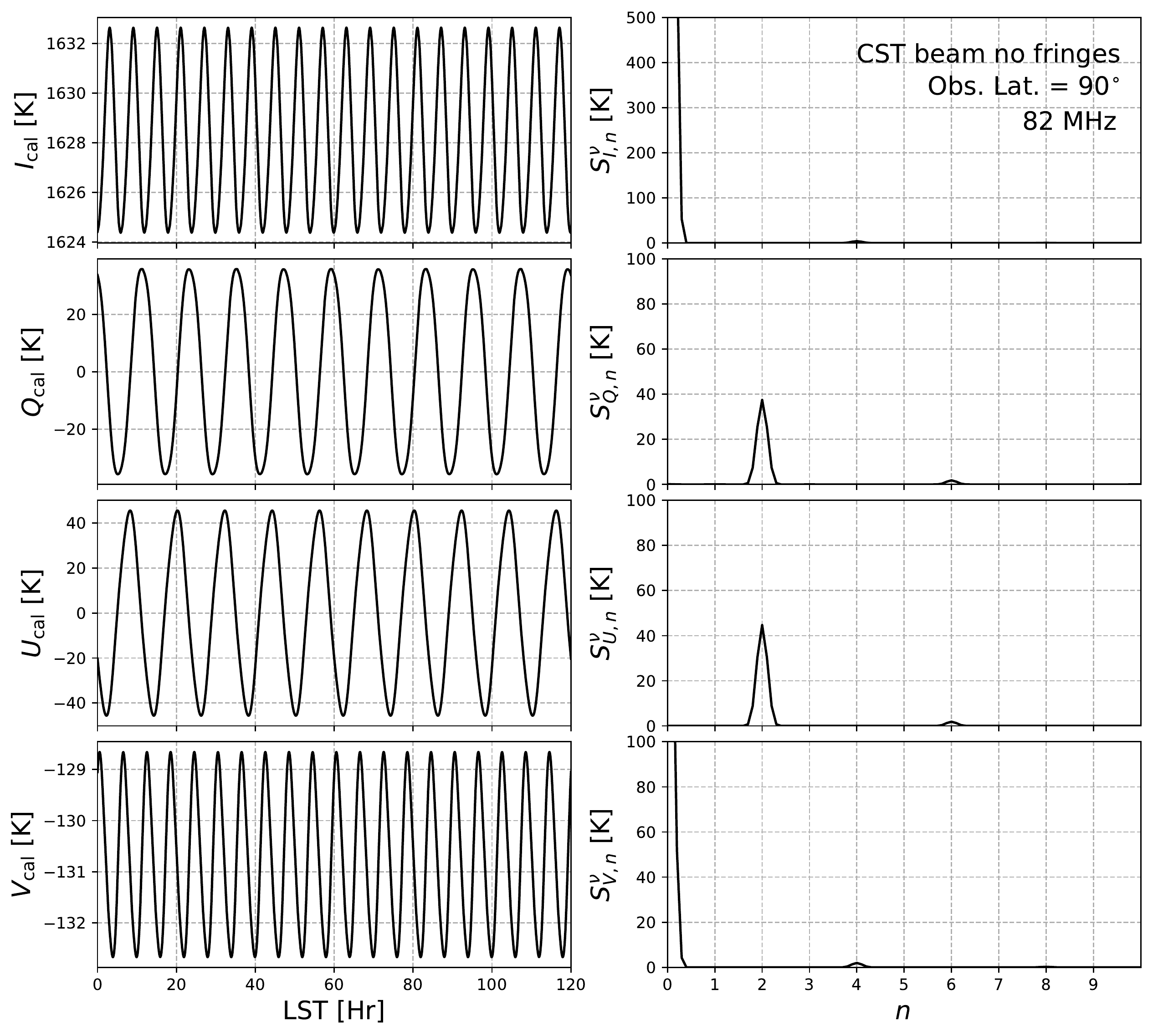}
  \caption{PIPE simulation result for the ground-parallel fiducial
    antenna beam at the GNP (Lat. $=\ 90^{\circ}$N). (Left) Temporal
    waveforms of the Stokes parameters (top to bottom: $I_{\rm cal},
    Q_{\rm cal}, U_{\rm cal}, V_{\rm cal}$) for multiple sidereal days
    (as shown here for 7 days and 14 cycles). (Right) Harmonic
    decomposition for the corresponding Stokes parameters on the left,
    note the strong twice-diurnal ($n=2$) component for Stokes $Q_{\rm
      cal}$ and $U_{\rm cal}$. Note that, although minute,
    imperfections on the CST beam give rise of the $n=4$ components
    for Stokes $I$ and $V$ as well as the $n=6$ components for Stokes
    $Q$ and $U$.}
  \label{fig:ctp_fft_ideal_beam_82MHz_north_pole_10day}
\end{figure}

Without loss of generality, by using this zenith-pointing beam at 82
MHz, the resulting net polarization from the PIPE simulation produces
sinusoidal waveforms with twice-diurnal period in Stokes $Q_{\rm cal}$
and $U_{\rm cal}$ in
\autoref{fig:ctp_fft_ideal_beam_82MHz_north_pole_10day}. Besides the
expected second harmonic ($n=2$), Fourier decomposition of these two
Stokes parameters for multiple consecutive sidereal days has also
identified a weak sixth harmonic ($n=6$). Additionally, the Stokes
$I_{\rm cal}$ and $V_{\rm cal}$, which should have been constant and
zero for a Gaussian beam, now contain a fourth harmonic ($n=4$). These
artifacts are due to deviations of the realistic CST beam from a
smooth Gaussian beam, which imply that Stokes parameters together can
provide a direct means to further characterize the beam systematics.

\subsection{Distortions from a Tilted Dipole}
\label{sec:ground_distortions}
With such a large beam (FWHM $\ge 60^{\circ}$) from the sleeved
dipole, potential interactions between the ground soil and the antenna
beam are expected, especially when the antenna had to be tilted at an
angle of $\delta_{\rm tilt} = \left(90^{\circ} - {\rm
  Observer\ Latitude}\right)$ relative to the horizontal ground to
point at the NCP.

\begin{figure}[!htb]
\centering
\includegraphics[width=0.95\columnwidth]{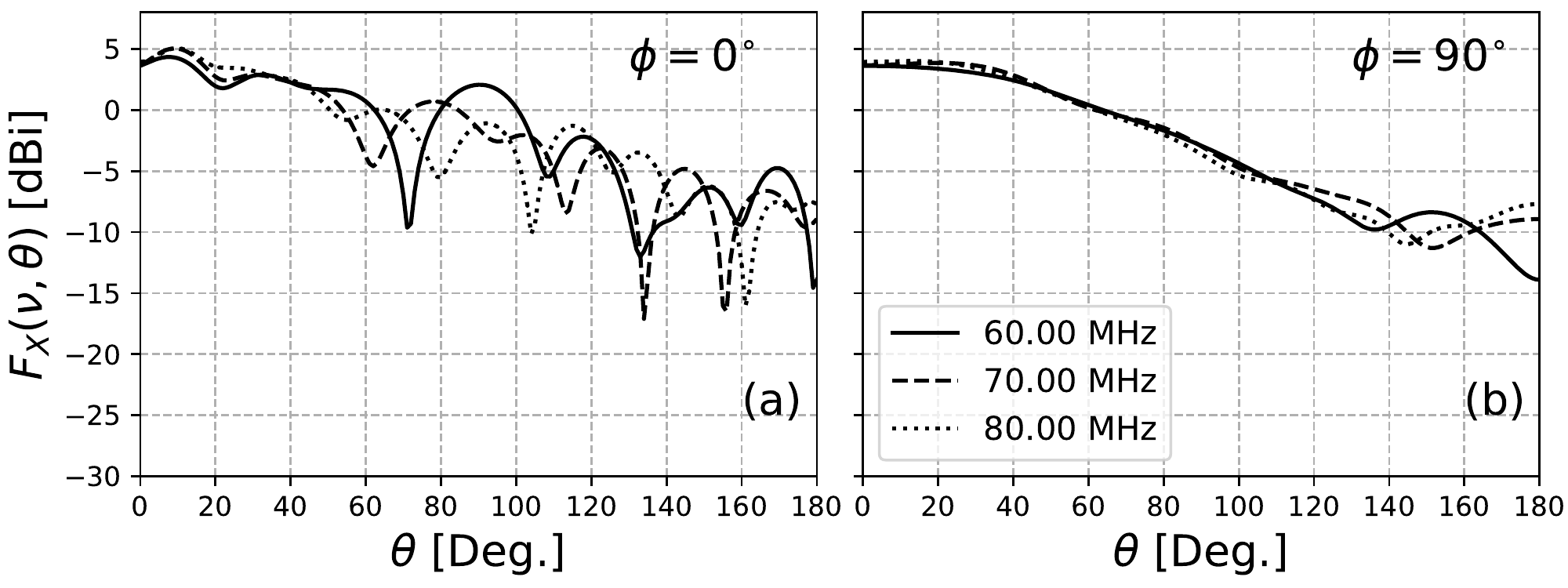}
\caption{Angular plots for the CST beam of the sleeved dipole's
  $X$-polarization when the antenna is tilted toward the NCP. As a
  result, the $E$- and $H$-planes of the beam are corrupted by the
  fringing structures due to interferometric interactions with the
  ground soil when the antenna is
  tilted.} \label{fig:cst_ctp_v1_40deg_x1200_y400_polX_slices_log_composite}
\end{figure}

\begin{figure}[!htb]
\centering 
\includegraphics[width=0.95\columnwidth]{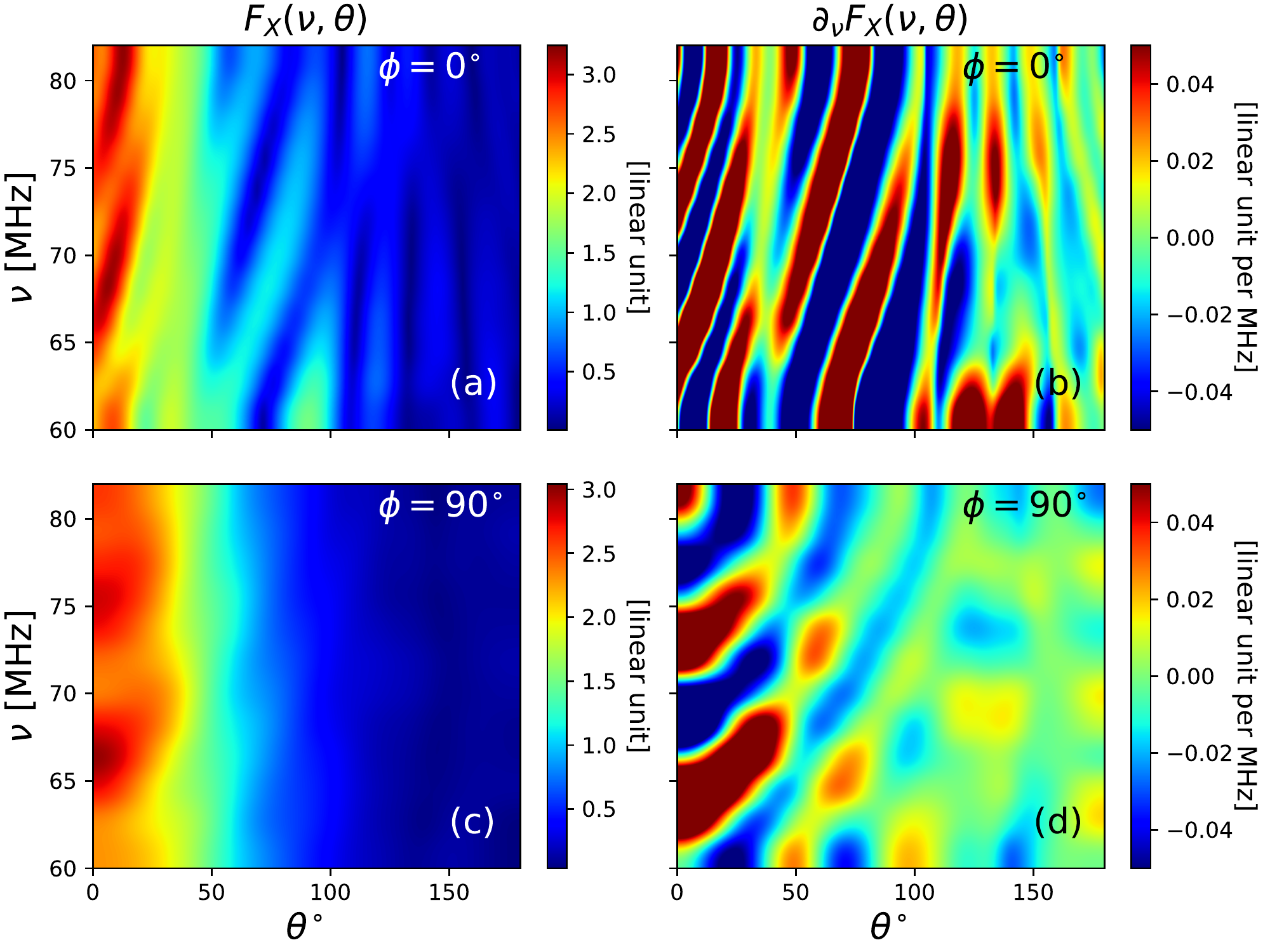}
\caption{(Left) 2D plots of the $E$- and $H$-planes of the CST beam
  model, $F_X(\tpn)$, when the sleeved dipole is tilted toward
  NCP. (Right) 2D plots of the frequency gradient,
  $\partial_{\nu}F_X(\tpn)$, of the beams on the left panels. The
  strong fringing structures due to interactions between the beam and
  ground soil are more apparent in the gradient plots, with
  $\left|\partial_{\nu}F(\tpn)\right|$ exceeds $
  0.05$.}\label{fig:cst_ctp_v1_40deg_x1200_y400_polX_gradient}
\end{figure}

As an example, the antenna model and its ground screen are tilted up
by $52^{\circ}$ relative to the ground soil slab in CST for an
observing latitude of $38^{\circ}$N. The resulting beams
(\autoref{fig:cst_ctp_v1_40deg_x1200_y400_polX_slices_log_composite}),
and their spectral gradients
(\autoref{fig:cst_ctp_v1_40deg_x1200_y400_polX_gradient}, right
column) show that strong fringes across the band, which are the
results of interferometric interactions between the antenna beam and
the ground soil, have corrupted the beam smoothness.

\begin{figure}
\centering 
\includegraphics[width=0.95\columnwidth]{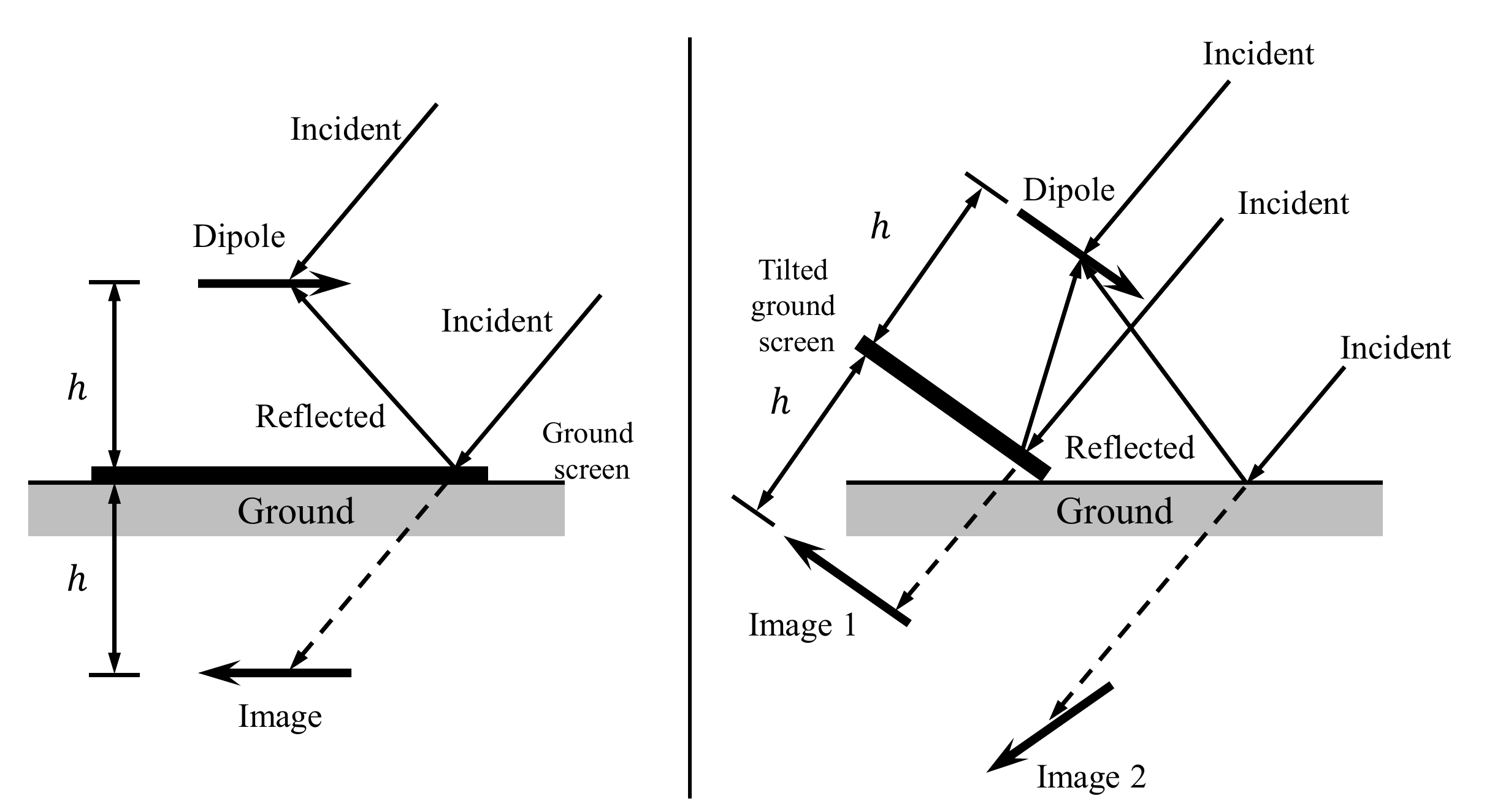}
\caption{Illustration comparing image theory for the two antenna
  configurations. (Left) Zenith-pointing horizontal dipole above
  ground screen at height $h$ produces a single image. (Right)
  Tilted dipole and ground screen relative to the ground soil can
  produce multiple images resulting in unwanted interferometric
  fringing which distorts the smooth beam pattern (Figure courtesy of
  NB17).}\label{fig:ground_image_comparison}
\end{figure}

This can be understood through image theory, as illustrated in
\autoref{fig:ground_image_comparison}. When a horizontal dipole
antenna locates above a finite ground screen at height $h$, it
produces a single image at the same distance under the ground
plane. However, when titling the antenna and its ground screen toward
the ground soil, the image below the soil and the one behind the
antenna's ground screen no longer overlap. The interferometric
interactions between these images subsequently imprint the unwanted
fringes onto the beams. In fact, this effect is dependent on the
antenna's directive gain since higher gain results in smaller
FOV. This effect can be mitigated by placing the antenna on a slope
with similar angle as the required tilting angle for
NCP-pointing. Suspending or mounting the tilted antenna far from the
ground can also help to reduce the fringing effects.

\subsection{Horizon Obstruction at Lower Latitudes}
\label{sec:horizon_obstruction}
\begin{figure}
\centering
\includegraphics[width=0.9\columnwidth]{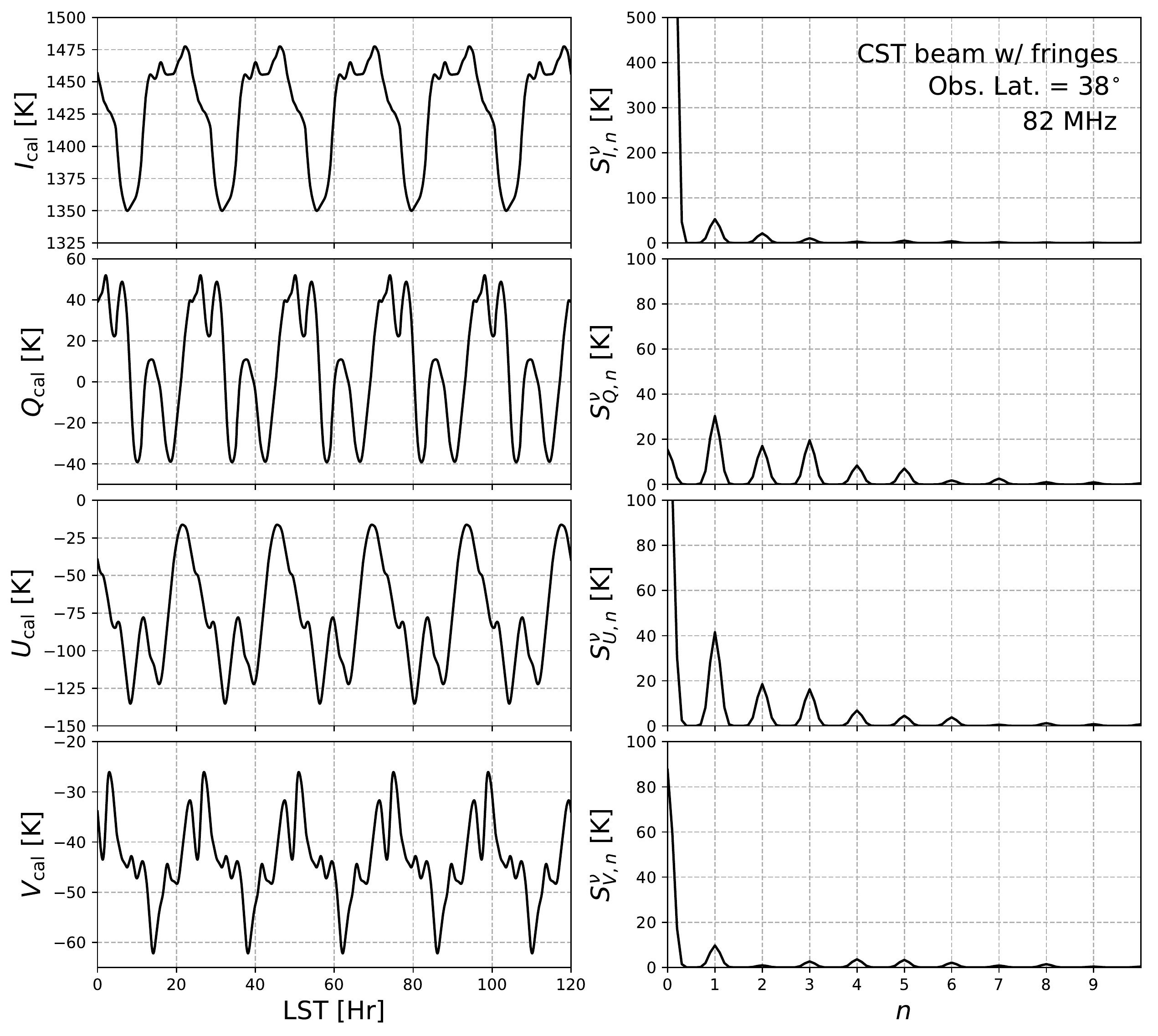}
\caption{PIPE simulation result for CST antenna beam at 82 MHz located
  at $38^{\circ}$N which produces the horizon cutoff of the northern
  sky. (Left) Waveforms of the Stokes parameters (top to bottom: $I,
  Q, U, V$) for multiple days of observation, note that the beam
  fringing effects and horizon cutoff have corrupted the symmetric
  sinusoidal waveforms from the idealized case. Instead they impose
  discontinuities and high-order harmonics to the waveforms. (Right)
  Fourier decomposition of the corresponding Stokes parameters from
  the left. The strong twice-diurnal components have been compromised
  by high-order terms.}
\label{fig:ctp_fft_40deg_beam_82MHz_tilted_farm_lat_10day}
\end{figure}

In addition to the beam distortions, at a latitude of $38^{\circ}$N,
the northern sky is partially obstructed by the Earth's horizon. The
continuously visible sky region over 24 sidereal hours is reduced when
sky rises and sets over the horizon. As a result, in combination with
distortions from the tilted beams, the smooth sinusoidal $Q_{\rm cal}$
and $U_{\rm cal}$ are replaced by waveforms with high-order harmonic
components
(\autoref{fig:ctp_fft_40deg_beam_82MHz_tilted_farm_lat_10day}). There
are still twice-diurnal components in Stokes $Q$ and $U$, but their
magnitude are reduced in the presence of the high-order terms. One
possible mitigation strategy is to relocate the instrument to a higher
latitude, closer to the geographic poles. This will help reducing the
tilting angle thus ground interactions with the beam as described in
the last section.

\section{The Cosmic Twilight Polarimeter}
\label{sec:ctp_experiment}
\subsection{Instrumentation and Data Acquisition}
\label{sec:ctp_instrumentation}
\begin{figure}
  \centering
  \includegraphics[width=0.8\columnwidth]{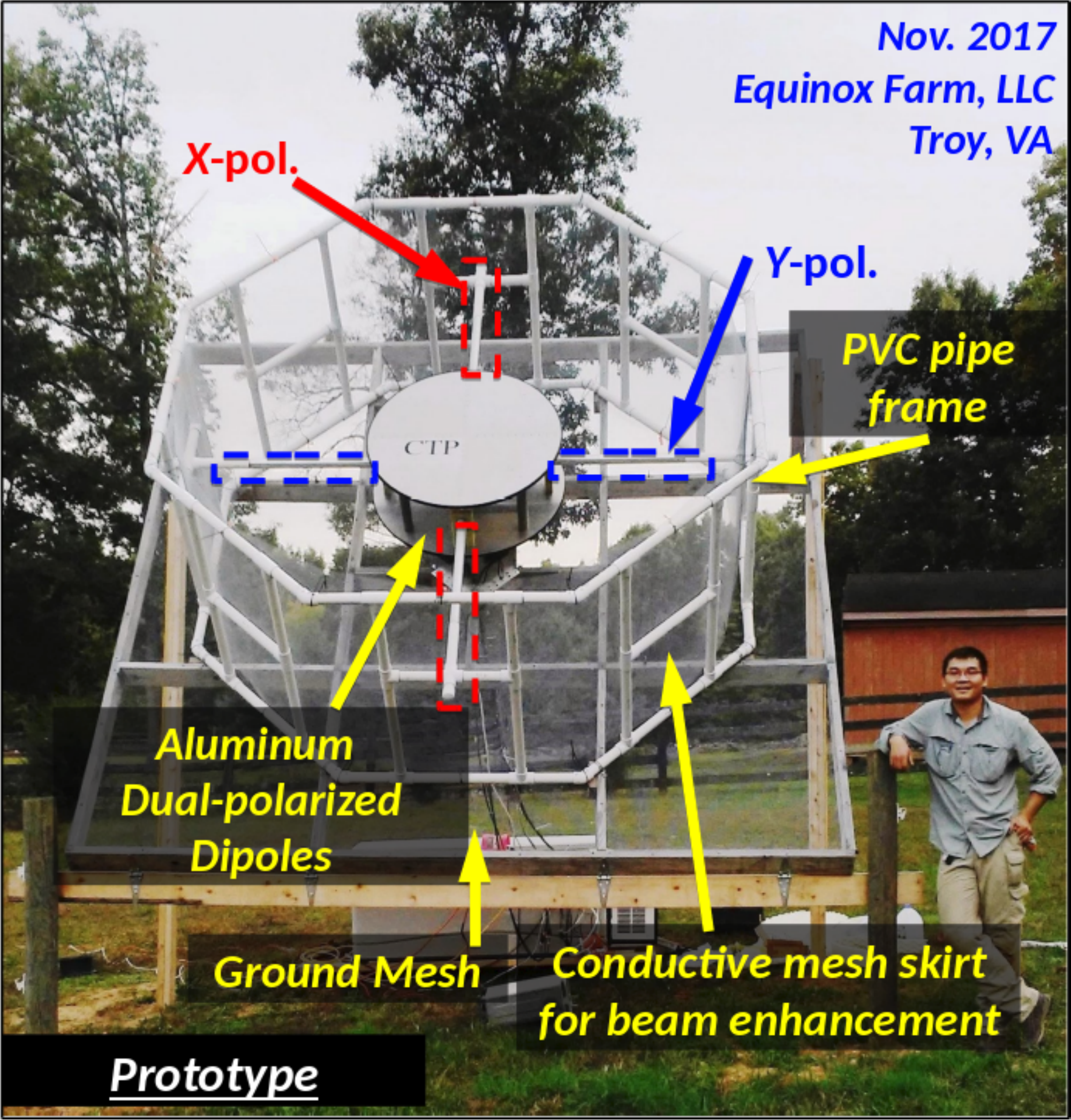} 
    \caption{Front view of the tilted sleeved dipole antenna pointing
      at the NCP at the deployment site in Troy, VA at latitude of
      $38.0^{\circ}$N, on a $\sim10^{\circ}$ slope. The antenna was
      mounted on top of a temperature controlled enclosure for the FE
      electronics. The antenna was configured such that the
      $Y$-polarization (blue) is horizontally parallel to the ground,
      and $X$-polarization (red) is tilted toward the
      ground.}\label{fig:ctp_field_deployment_illustration}
\end{figure}

The CTP adopted the sleeved dipole antenna design described in the
previous section. The system electronics consisted of a thermally
stabilized front-end (FE) stage along with a back-end (BE) instrument
rack stored in a weatherproof and thermally regulated enclosure. The
prototype was deployed during the Fall of 2017 at the Equinox Farm,
LLC, in Troy, VA ($38.0^{\circ}$N, $78.3^{\circ}$W) as a testbed for
the system integration, calibration, and preliminary measurement of
the PIPE. The CTP used the Polaris during night time to align the
pointing to the NCP. The general layout of the CTP is illustrated in
\autoref{fig:ctp_field_deployment_illustration}. Due to logistical
constraints, we were only able to place the CTP on a shallow
north-facing slope ($\sim 10^{\circ}$) in an attempt to alleviate some
of the ground interactions. The final tilting angle is about
$42^{\circ}$ from horizon.

The signal from each polarization was amplified by a low-noise
amplifier (LNA) and filtered through a radio-frequency (RF) module in
the FE. Although the sleeved antenna was designed to operate between
60-120 MHz, a 30-MHz bandpass filter (BPF) centered at 75 MHz was used
to reject radio frequency interference (RFI) from local digital TV
stations and the FM band (88-108 MHz).

\begin{figure}
\centering
\includegraphics[width=0.7\columnwidth]{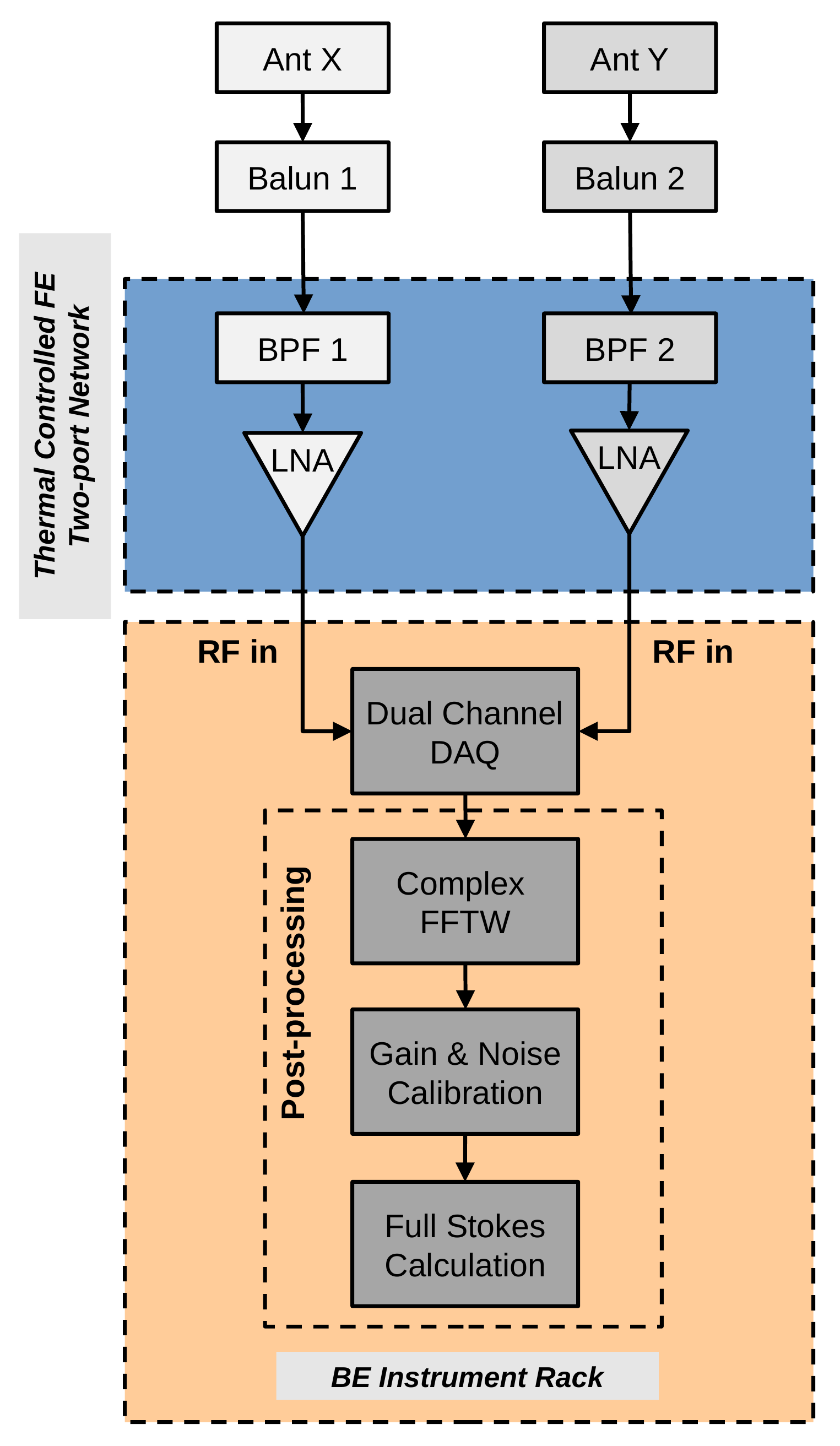}
\caption{Block diagram for the CTP's instrument layout. The instrument
  FE (blue shaded) consists of a the thermal controlled stage for the
  main RF signal chain. The BE instrument rack consists of a
  FPGA-based signal digitizer. The sampled signal are channelized into
  spectrum with the FFTW program before the signal gain and noise
  temperature are corrected. In the end, Stokes parameters are
  calculated and monitored as a function of time before applying the
  harmonic decomposition on them for further
  analysis.}\label{fig:daq_pipeline_simplified}
\end{figure}

The output voltages from both polarizations were digitized by a
FPGA-based analog-to-digital converter (ADC), Signatec PX14400A,
before being channelized with fast Fourier transform (FFT), using the
FFTW3\footnote{\texttt{http://www.fftw.org}, \cite{frigo2005design}}
software library, into complex voltages $\widetilde{V}_X(t,\nu)$ and
$\widetilde{V}_Y(t,\nu)$ at a resolution bandwidth (RBW) of
$\Delta\nu\sim57.00$ kHz. A block diagram for the instrument layout is
provided in \autoref{fig:daq_pipeline_simplified}.

Subsequently, the autocorrelation
($\langle\widetilde{V}_X\widetilde{V}_X^*\rangle$,
$\langle\widetilde{V}_Y\widetilde{V}_Y^*\rangle$) and
cross-correlation ($\langle\widetilde{V}_X\widetilde{V}_Y^*\rangle$,
$\langle\widetilde{V}_Y\widetilde{V}_X^*\rangle$) were calculated and
time averaged. The resulting uncalibrated net Stokes parameters were
computed as
\begin{align}
  I_{\rm uncal}(t,\nu) &= \langle \widetilde{V}_X\widetilde{V}_X^*\rangle +
  \langle \widetilde{V}_Y\widetilde{V}_Y^*\rangle, \label{eq:stokes_I_uncal} \\
  Q_{\rm uncal}(t,\nu) &= \langle \widetilde{V}_X\widetilde{V}_X^*\rangle -
  \langle \widetilde{V}_Y\widetilde{V}_Y^*\rangle, \label{eq:stokes_Q_uncal}\\
  U_{\rm uncal}(t,\nu) &= \langle \widetilde{V}_X\widetilde{V}_Y^*\rangle +
  \langle \widetilde{V}_X^*\widetilde{V}_Y\rangle, \label{eq:stokes_U_uncal}\\
  V_{\rm uncal}(t,\nu) &= i\left(\langle \widetilde{V}_X\widetilde{V}_Y^*\rangle -
  \langle \widetilde{V}_X^*\widetilde{V}_Y\rangle\right). \label{eq:stokes_V_uncal}
\end{align}
The Stokes parameters were converted to temperature units after
calibrating for the multiplicative transducer gain, $G_{T}(t,\nu)$,
and the additive instrumental noise temperature, $T_n(t,\nu)$, through
the use of the network-theory based calibration equations with
\begin{align}
  \begin{split}
    I_{\rm cal}(t,\nu) ={} &\frac{1}{k_B\Delta\nu}\left[ \left(\frac{\langle
        \widetilde{V}_X\widetilde{V}_X^*\rangle}{G_{T,X}} + \frac{\langle
        \widetilde{V}_Y\widetilde{V}_Y^*\rangle}{G_{T,Y}}\right) - \right.\\
      &\left. \left(T_{n,X}+T_{n,Y}\right)\vphantom{\frac{\langle
          \widetilde{V}_X\widetilde{V}_X^*\rangle}{G_{T,X}}}\right], \label{eq:stokes_I_cal}
  \end{split}\\
  \begin{split}    
    Q_{\rm cal}(t,\nu) ={} &\frac{1}{k_B\Delta\nu}\left[ \left(\frac{\langle
        \widetilde{V}_X\widetilde{V}_X^*\rangle}{G_{T,X}} - \frac{\langle
        \widetilde{V}_Y\widetilde{V}_Y^*\rangle}{G_{T,Y}}\right) - \right.\\
      &\left. \left(T_{n,X}-T_{n,Y}\right)\vphantom{\frac{\langle
          \widetilde{V}_X\widetilde{V}_X^*\rangle}{G_{T,X}}}\right], \label{eq:stokes_Q_cal}
  \end{split}\\
  \begin{split} 
    U_{\rm cal}(t,\nu) ={} &\frac{2}{k_B\Delta\nu} \frac{{\rm Re}\left(\langle
      \widetilde{V}_X\widetilde{V}_Y^*\rangle\right)}{\sqrt{G_{T,X}G_{T,Y}}}, \label{eq:stokes_U_cal}
  \end{split}\\
  \begin{split} 
    V_{\rm cal}(t,\nu) ={} &\frac{-2}{k_B\Delta\nu} \frac{{\rm Im}\left(\langle
      \widetilde{V}_X\widetilde{V}_Y^*\rangle\right)}{\sqrt{G_{T,X}G_{T,Y}}}, \label{eq:stokes_V_cal}
  \end{split}
\end{align}
where the subscripts in $G_T(t,\nu)$ and $T_n(t,\nu)$ refer to
polarizations $X$ and $Y$. The derivation of the calibration equations
are provided in \autoref{appdx:stokes_calibration} and the
  procedures to determine $G_T(t,\nu)$ and $T_n(t,\nu)$ are described
  in the following section.

\subsection{Network-theory Based Calibration}
\label{sec:network_calibration}
In a conventional total-power experiment, one of the primary purposes
of the calibration is to remove the multiplicative gain and additive
noise temperature of the system to recover the apparent antenna
temperature from the measured power. One of the simplest calibration
schemes to implement is to correct the power gain and noise
temperature for the sky signal power based on the measured power of a
reference load. A Dicke radiometer calibration is one example based on
such on-off reference load scheme \citep{dicke1982measurement}. More
sophisticated reference-load calibration schemes also rely on
continuously recording and comparing the antenna power to the ones
from multiple reference loads or a broadband noise source, such as the
variants adopted by EDGES \citep{rogers2012absolute} and SARAS
\citep{patra2013saras}.

\begin{figure}
\centering 
\includegraphics[width=\columnwidth]{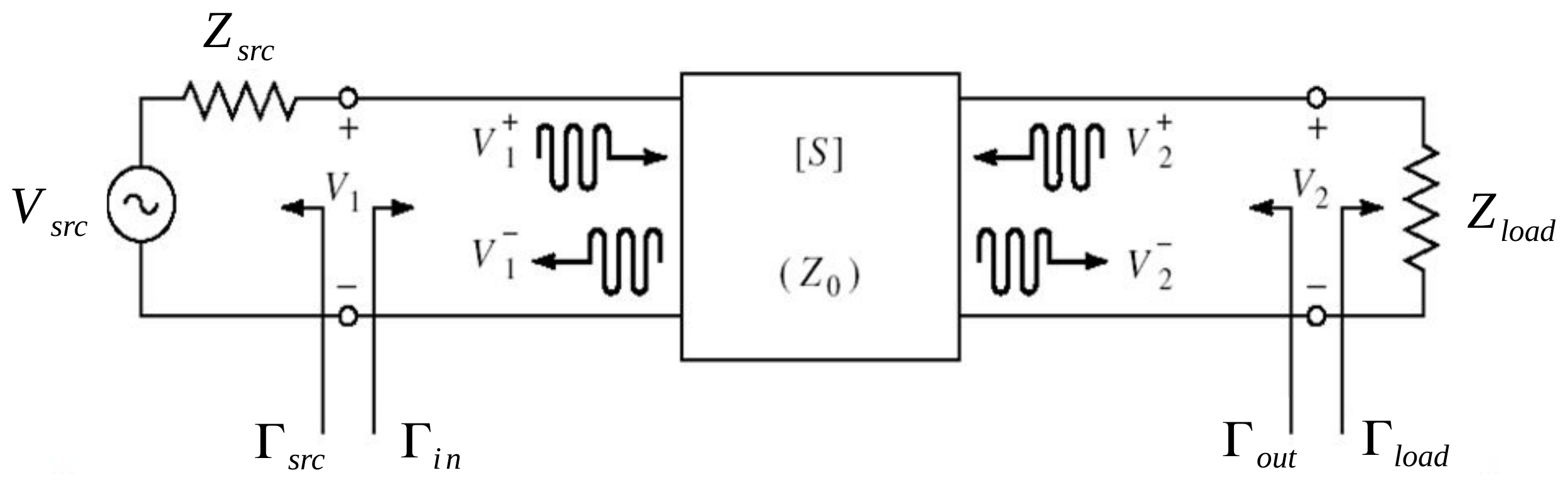}
\caption{Diagram for a generic two-port network, with input source
  impedance $Z_{\rm src}(\nu)$ and output load impedance $Z_{\rm
    load}(\nu)$. The incident ($V^{+}$) and reflected ($V^{-}$)
  voltages at the input and output ports are related by the scattering
  matrix $\bm{S}(\nu)$ at some network characteristic impedance
  $Z_0$. $\Gamma$ is the reflection coefficients due to the
  corresponding impedance at the given direction of the
  arrow.}\label{fig:two_port_network}
\end{figure}

In general, these calibration schemes assume the power gain and noise
temperature of the system to be constant when toggling between the
antenna and different reference calibrators over a short time
interval. However, according to electrical network theory
\citep{collin2007foundations, engberg1995noise}, a
linear\footnote{This is assumed if the device operates in the linear
  regime. For example, an active device like an LNA is not overdriven,
  which results in gain compression into the nonlinear regime.}
two-port network (\autoref{fig:two_port_network}), its power gain and
noise temperature depend on the impedance of other devices connected
at the network's input and output ports. Namely, the gain and noise
temperature measured between the antenna and loads are not
representative of one another since their impedance are not
necessarily identical. Hence, determining the gain and noise
temperature directly as a function of frequency can pose a
challenge. For these reasons, a network-theory based calibration
approach was adopted for the CTP. This scheme relies on
impedance-independent network parameters to determine the power gain
and noise temperature.

\subsubsection{Transducer Gain Correction with $S$-Parameters}
\label{sec:gain_calibration}
For a linear two-port network, the power gain can be described by the
transducer gain $G_T(\nu)$ as \citep{collin2007foundations},
\begin{equation}
    G_T(\nu)  =\frac{\left(1-|\Gamma_{\rm src}|^2\right)\left(1-|\Gamma_{\rm
          load}|^2\right)|S_{21}|^2}{|\left(1-S_{11}\Gamma_{\rm src}\right)
      \left(1-S_{22}\Gamma_{\rm
          load}\right)-S_{12}S_{21}\Gamma_{\rm src}\Gamma_{\rm load}|^2},
    \label{eq:transducer_gain_def}
\end{equation}
where $S_{11}(\nu)$, $S_{12}(\nu)$, $S_{21}(\nu)$, and $S_{22}(\nu)$
are the complex components of the scattering matrix
$\bm{S}(\nu)$. This set of $S$-parameters describes the intrinsic
properties of the device under test (DUT) and can be measured directly
with a Vector Network Analyzer (VNA) in the laboratory. Meanwhile,
$\Gamma_{\rm src}(\nu)$ is the complex reflection coefficient of the
DUT arising from impedance mismatch between the input source and the
DUT's input port. Similarly, $\Gamma_{\rm load}(\nu)$ is the
reflection coefficient caused by the mismatch between the network's
output and a load\footnote{The system load is distinct from the
  reference loads at the input port of the DUT. By definition, the
  load impedance refers to the one of any device connected at the
  output port of the device.} connected to the output port. The
reflection coefficient is defined as $\Gamma_i = (Z_i-Z_0)/(Z_i+Z_0)$,
where $Z_0$ is the network's characteristic impedance.

For CTP, since the entire signal chain of each polarization (from the
input of the FE modules through the output at the end of the coaxial
cables before entering the digitizer) was considered as a single
two-port network, end-to-end VNA measurements were made for the
$S$-parameters. As a result, the $S$-parameters had also included the
ohmic loss in the coaxial cables and other passive RF devices along
the signal path before entering the digitizer.

It is worth pointing out that the $G_T$ defined in Equation
\eqref{eq:transducer_gain_def} has included the reflection efficiency
terms due to impedance mismatch between the antenna and receiver as
the first two terms in the numerator, i.e., $\eta_{\rm tot}(\nu) =
\eta_{\rm ant}(\nu)\eta_{\rm rcv}(\nu) = \eta_{\rm src}(\nu)\eta_{\rm
  load}(\nu)= \left(1-|\Gamma_{\rm ant}|^2\right)\left(1-|\Gamma_{\rm
  rcv}|^2\right)$ if the input source is the antenna and the output
load is the receiver. On CTP, since the DUT's load is the digitizer
whose input ports' impedance are well matched to the CTP's output
($\sim 50\ \Omega$), we assumed $\eta_{\rm load} \sim 1$.
  
Furthermore, the $S$-parameters are functions of operating temperature
since the electronic components are susceptible to thermal
variations. To mitigate such fluctuations during observation, the CTP
was equipped with an active thermal control system at the FE
enclosure, using a thermoeletric Peltier cooler powered by a
proportional-integral-derivative (PID) feedback circuit. Additionally,
the $S$-parameters of the FE modules were measured with a VNA when
operating at different set temperatures in a laboratory thermal
enclosure, over a range of $\sim 20$-$35^{\circ}$C.

These measurements helped to establish a thermal dependence for the
$S$-parameters, which was used to correct for low-level temperature
variations insensitive to the PID controller. Subsequently, a set of
calibration coefficients were acquired by least-squares fitting these
$S$-parameter measurements as functions of temperature and
frequency. These coefficients helped to interpolate the $S$-parameters
to an in situ ambient temperature $T_{\rm amb}$ recorded during each
observation instance, which in return determined $G_T(\nu,T_{\rm
  amb})$ when $\Gamma_{\rm src}$ is substituted by $\Gamma_{\rm ant}$
measured in the field. Dependence of $G_T(\nu,T_{\rm amb})$ as a
function of $T_{\rm amb}$ is shown in
\autoref{fig:trans_gain_correction} in
\autoref{appdx:lab_calibration}.

\subsubsection{Noise Temperature Correction with Noise Parameters}
\label{sec:noise_calibration}
Analogous to $G_T(\nu)$, the noise temperature, $T_n(\nu)$, is also a
function of the input source impedance $Z_{\rm src}(\nu)$. It can be
parametrized in terms of a set of four noise parameters intrinsic to
the DUT, like the $S$-parameters. According to noise theory,
$T_n(\nu)$ reaches a minimum value at $T_{\rm min}(\nu)$ when the
source impedance is matched to an optimal value $Z_{\rm opt}(\nu)$ as
\citep{engberg1995noise},
\begin{equation}
T_n(\nu) = T_{\rm min} + \frac{4NT_0|\Gamma_{\rm src} - \Gamma_{\rm
    opt}|^2}{|1 + \Gamma_{\rm opt}|^2\left(1 - |\Gamma_{\rm
    src}|^2\right)},
\label{eq:noise_parameter}
\end{equation}
where the dimensionless $N(\nu) = R_n(\nu)G_{\rm opt}(\nu)$ is the
product of the equivalent noise resistance $R_n(\nu) = \langle
v_n^2\rangle/(4k_BT_0\Delta\nu)$ and the optimal source conductance
$G_{\rm opt}(\nu)$, with $\langle v_n^2\rangle$ is the mean-square
noise-generator voltage and $T_0 = 290$ K\footnote{The standard noise
  temperature is conventionally defined to be 290 K so that $k_BT_0
  \approx 4.00 \times 10^{-21}$ Ws.}. The complex source reflection
coefficient and optimal impedance are defined as $\Gamma_{\rm
  src}(\nu)$ and $\Gamma_{\rm opt}(\nu)$ respectively. Including the
real and imaginary parts of $\Gamma_{\rm opt}(\nu)$, the set of noise
parameters needed to determine $T_n(\nu)$ for the CTP system is
$\Big\{T_{\rm min}(\nu),\ {\rm Re}[\Gamma_{\rm opt}(\nu)],\ {\rm
  Im}[\Gamma_{\rm opt}(\nu)],\ N(\nu)\Big\}$.

To determine the DUT's noise parameters, a set of $T_n(\nu)$ were
measured in the laboratory with a set of reference input fixtures
consisting of simple passive electronic components ($50\ \Omega$,
$75\ \Omega$, $100\ \Omega$, $RC$-circuit, $RL$-circuit). These
fixtures provide source reflection coefficients more simple than the
$\Gamma_{\rm ant}$ thus simpler frequency structures in
$T_n(\nu)$. The four noise parameters were simultaneously fitted for
the measured $T_n(\nu)$ with the corresponding $\Gamma_{\rm src}(\nu)$
using the Python implementation of the Markov chain Monte Carlo (MCMC)
sampler,
\texttt{emcee}\footnote{\texttt{http://dfm.io/emcee/current/}}.
Combining with the field-measured $\Gamma_{\rm ant}(\nu)$, these
fitted noise parameters help to determine $T_n(\nu)$ for the
observational data. Details of this procedure and fitted noise
parameters from different $T_n(\nu,Z_{\rm src})$ are illustrated in
\autoref{appdx:lab_calibration}.

\begin{figure}
\centering
\includegraphics[width=\columnwidth]{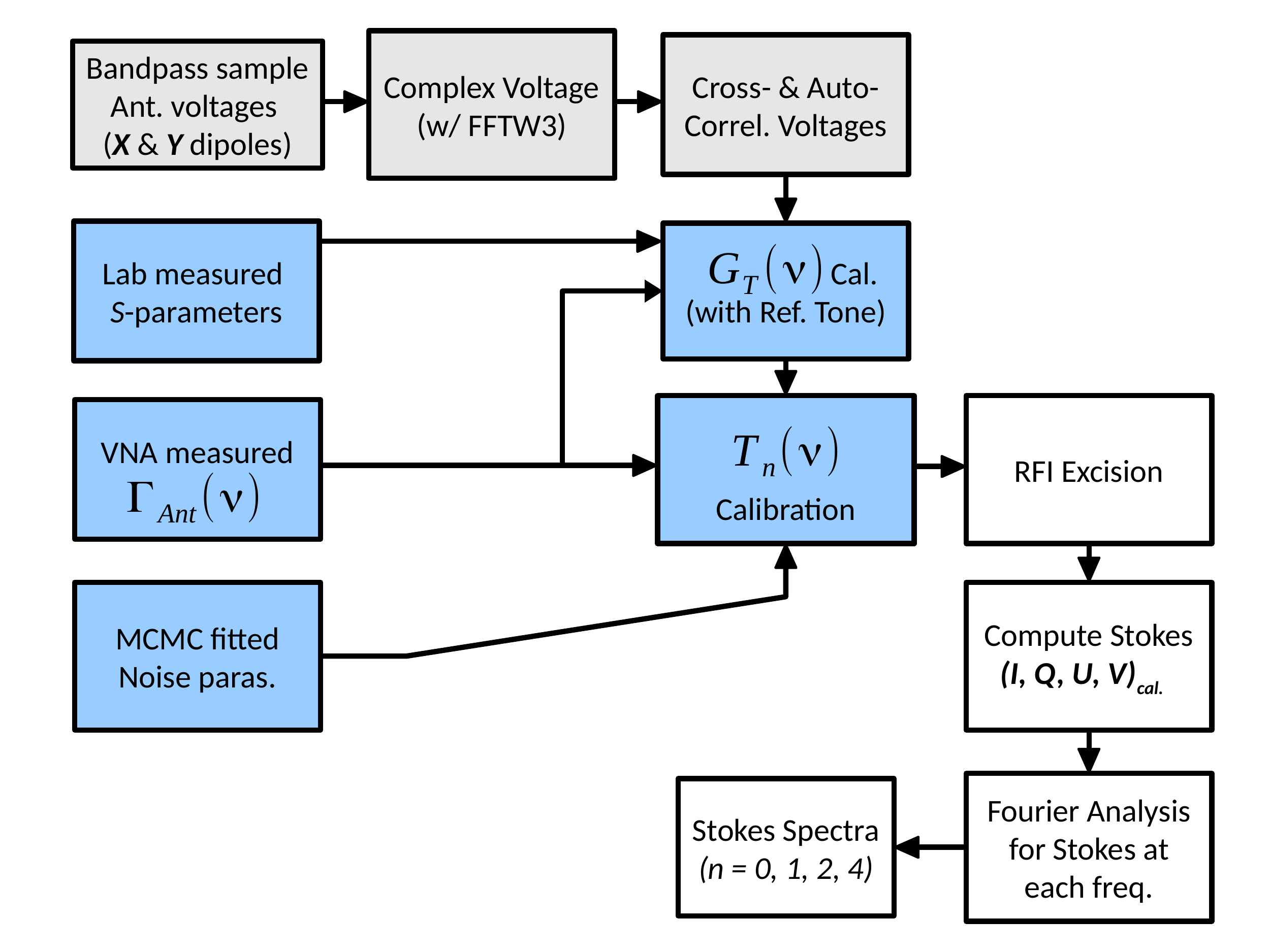}
\caption{Block diagram for the data pipeline for acquisition (gray
  shaded boxes), calibration (blue shaded) and reduction
  (white).}\label{fig:ctp_pipeline_block_diagram_nostepping}
\end{figure}

\subsection{Preliminary Polarization Constraints}
\label{sec:prelim_pol_constraint}
Since the CTP was deployed outside the radio-quiet zone, the observed
spectra were expected to be contaminated by RFI. Although the BPF
between 60-90 MHz has substantially reduced the RFI from the FM band,
majority of the band was still corrupted. Nonetheless, a small subband
around 82 MHz was determined to be the cleanest for the preliminary
analysis.

As summarized in \autoref{fig:ctp_pipeline_block_diagram_nostepping},
after applying the aforementioned calibration for $G_T(\nu)$ and
$T_n(\nu)$, kurtosis-based RFI excision was applied to the auto- and
cross-correlated voltages for the two polarizations before computing
the Stokes parameters\footnote{To constrain the PIPE in the tilting
  configuration, no beam correction were applied when comparing the
  observed Stokes with simulation. However, to properly correct the
  beam pattern, CST simulation will be needed. The CST beams have
  included the ohmic loss when proper materials are assigned to the
  detailed antenna model. The antenna ohmic loss can be constrained
  through the ohmic loss efficiency $\epsilon_r(\nu)$ provided in CST,
  where typically the lossy antenna effective area is defined as
  $A_e({\rm lossy}) \sim \epsilon_r A_e({\rm lossless}$).}  FFTs were
then applied to the four Stokes parameters to extract the harmonic
components. In order to resolve any underlying harmonics, the time
data length needs to be increased. In practice, multiple consecutive
days of data will be concatenated into single data streams prepared
for the FFT.

Within the 10 days of observation, the 24 hours of data collected on
Dec $1^{\rm st}$, 2017 are the least contaminated by RFI and other
spurious distortions. To achieve the needed resolution for the
low-order harmonics in the Stokes PSD, that single day of data were
duplicated for 10 times to emulate 10 days of continuous
observation. Additionally, to suppress some of the high-frequency
noise in the observed data, a Savitzky-Golay filter was applied to
concatenated data streams as a moving mean before computing the FFT.

The window size of the filter was chosen to be around $2\%$ of the
total data length. Since the larger the window size is, the more
high-order harmonics are filtered out. After comparing different
window sizes, the $2\%$ window width was determined to provide a good
balance between preserving the fidelity of the lowest four harmonics
($n \ge 4$) and rejecting some of the noise as well as discontinuities
due to RFI excision. The window-size optimization process is
illustrated in
\autoref{fig:ctp_obs_smth_win_sz_compare_82MHz_1nday_doy335} in
\autoref{sec:savgol_win_sz_effect}.

The PIPE simulation suggests that no significant harmonics should
appear for $n \geq 10$. The confidence intervals of the harmonics are
standard deviations of the Stokes PSD for $n \geq 10$ before applying
the Savitzky-Golay filter. They are computed as,
\begin{equation}
  \hat{\sigma}_{S_i}^2 = {\frac{1}{(n_{\rm max}-n_{\rm
        min} + 1)}\sum_{n=n_{\rm min}}^{n_{\rm max}}\left(S_{S_i,n}^{\nu} - \hat{\mu}_{S_i}^{\nu}\right)^2},
  \label{eq:std_noise}
\end{equation}
where the estimated sample mean of the noise
floor in the Stokes PSD is determined by,
\begin{equation}
  \hat{\mu}_{S_i}^{\nu} = {\frac{1}{(n_{\rm max}-n_{\rm
        min})}\sum_{n=n_{\rm min}}^{n_{\rm max}}S_{S_i,n}^{\nu}},
  \label{eq:mu_noise_floor}
\end{equation}
with $[n_{\rm min}, n_{\rm max}] = [10,N]$.

\begin{figure}
\centering
\includegraphics[width=.85\columnwidth]{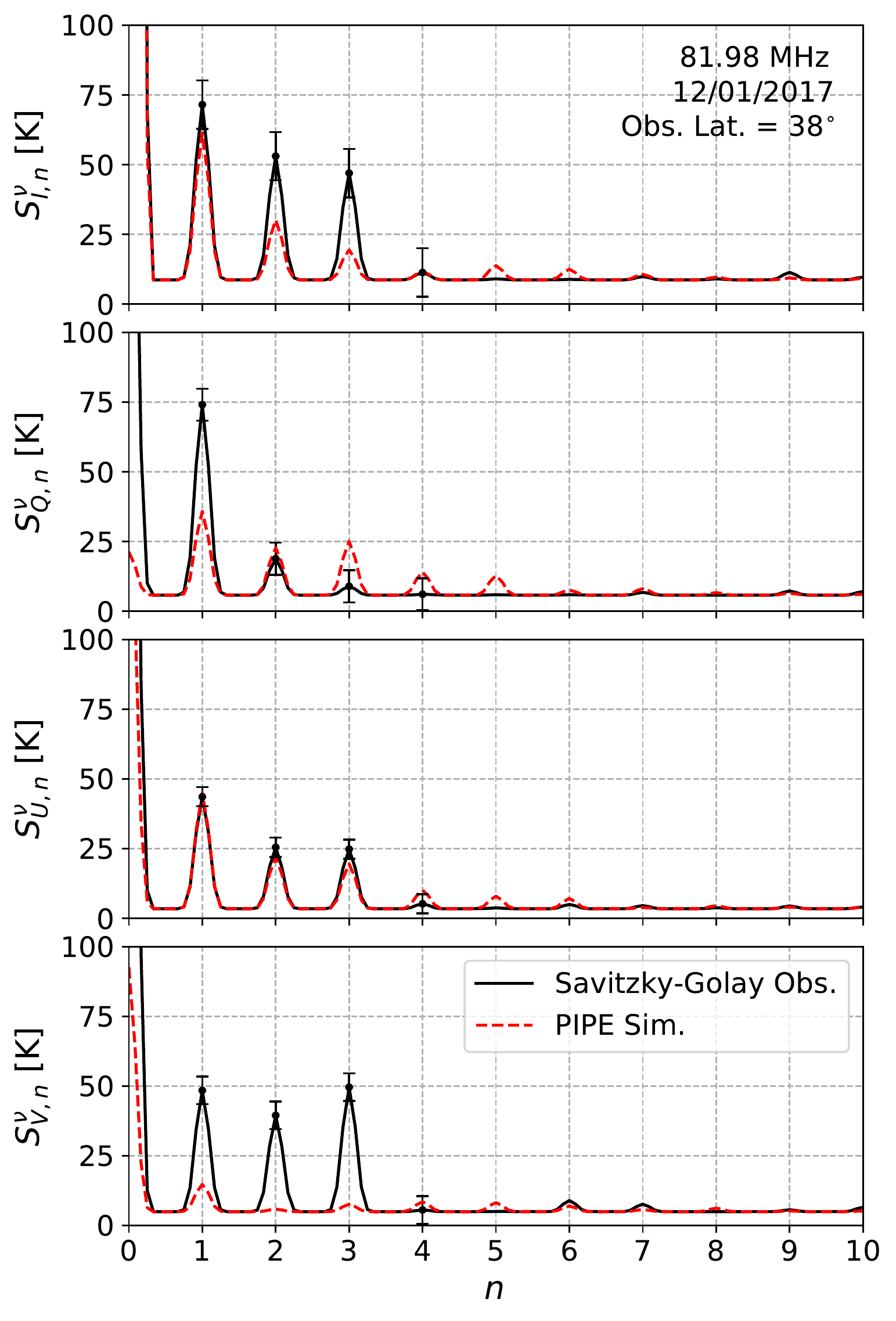}
\caption{Comparison of the harmonic decomposition, up to $n = 10$,
  between the observation (black curve) and PIPE simulation with the
  tilted beam and horizon cutoff (red curve) at $\sim 82$
  MHz. To better compare PSD of the Stokes waveforms
  smoothed by the Savitzky-Golay filter, the observation noise
  $\hat{\sigma}_{S_i}$ was added to the noiseless PIPE simulation in
  the analysis. $S_{Q,n}^{\nu}$ and $S_{U,n}^{\nu}$ strongly indicate there is
  induced polarization observed in the data. However, due to the
  limited number of clean channels, further evaluation is needed when
  the CTP is redeployed to a more RFI-quiet environment. Note that
  there is a slight shift in the observed harmonics comparing to the
  simulation, this is determined to be inaccuracy in cadence when
  concatenating multiple days of data into a single data stream for
  the FFT computation.}\label{fig:ctp_sim_fft_compare_82MHz_1nday_doy335}
\end{figure}

Magnitudes of the identified harmonics at $n = \{1,\ 2,\ 3,\ 4\}$ and
corresponding uncertainties $\hat{\sigma}_{S_i}$ in the Stokes
waveforms are listed in \autoref{tbl:harmonic_ratio} and shown in
\autoref{fig:ctp_sim_fft_compare_82MHz_1nday_doy335}. From the
observed data, the signal-to-noise ratio (S/N) at the twice-diurnal
components in Stokes $Q$ and $U$ are computed to be 3.27 and 7.40,
respectively.

\begin{table}
  \centering
  \caption{Observed and Simulated Stokes harmonics at 81.98
    MHz}\label{tbl:harmonic_ratio} \centering
  \begin{tabular}{c c c c c}
    \hline
    \hline
    \T\B  & $S_{i,1}$ [K] & $S_{i,2}$ [K] & $S_{i,3}$ [K] & $S_{i,4}$ [K] \\
    \hline
    \T Obs. &  &  & & \\
    $I$ & 71.53 $\pm\ $8.66 & 53.05 $\pm\ $8.66 & 46.98 $\pm\ $8.66 & 11.33 $\pm\ $8.66 \\
    $Q$ & 74.08 $\pm\ $5.75 & 18.85 $\pm\ $5.75 & 8.95 $\pm\ $5.75 & 6.10 $\pm\ $5.75 \\
    $U$ & 43.59 $\pm\ $3.44 & 25.46 $\pm\ $3.44 & 24.81 $\pm\ $3.44 & 5.22 $\pm\ $3.44 \\ 
    \B $V$ & 48.42 $\pm\ $4.96 & 39.47 $\pm\ $4.96 & 49.57 $\pm\ $4.96 & 5.57 $\pm\ $4.96 \\
    \hline
    \T Sim. \\
    $I$ & 61.38 & 30.23 & 19.44 & 11.72 \\
    $Q$ & 35.84 & 22.73 & 25.08 & 14.09 \\
    $U$ & 45.01 & 21.86 & 19.51 & 10.12 \\
    \B$V$ & 14.75 & 5.84 & 7.59 & 8.50 \\
    \hline
    \T Ratio & Obs./Sim.\\
    $I$ & 1.17 $\pm\ $0.22 & 1.75 $\pm\ $0.58 & 2.42 $\pm\ $1.17 & 0.97 $\pm\ $1.03 \\
    $Q$ & 2.07 $\pm\ $0.37 & 0.83 $\pm\ $0.33 & 0.36 $\pm\ $0.24 & 0.43 $\pm\ $0.44 \\
    $U$ & 0.97 $\pm\ $0.11 & 1.16 $\pm\ $0.24 & 1.27 $\pm\ $0.29 & 0.52 $\pm\ $0.38 \\
    \B$V$ & 3.28 $\pm\ $1.15 & 6.76 $\pm\ $5.80 & 6.53 $\pm\ $4.32 & 0.66 $\pm\ $0.70 \\
    \hline
  \end{tabular}
\end{table}

Relative magnitude ratios between the observed low-order harmonics and
ones from PIPE simulation at 81.98 MHz are also presented in
\autoref{tbl:harmonic_ratio}. The observed harmonics for
$n=\{1,\ 2\ ,3\}$ in Stokes $U$ are the most consistent to the
simulation (with their ratios close to unity within
uncertainty). Meanwhile $S_{I,1}$ is also consistent with the
simulation since this represents the bulk part of the diurnal
component when the Galaxy rises and sets. However, Stokes $Q$ and $V$
are stronger in the observation.

These discrepancies most likely arose from the observed stronger
signal measured in polarization $Y$ (horizontally oriented) comparing
to $X$ (\autoref{fig:ctp_field_deployment_illustration}). Additional
signal could also have been reflected and scattered off the ground
when the antenna was tilted forward. Since Stokes $Q$ was computed as
the difference between the autocorrelated power of $X$ and $Y$ in
Equation \eqref{eq:stokes_Q_uncal}, uneven signal power measured
between dipole $X$ and $Y$ could introduce an offset to Stokes
$Q$. Some of the scattered signal might have also contributed to the
circular polarization as seen in the stronger Stokes $V$ harmonics
(\autoref{fig:ctp_field_deployment_illustration}). Follow-up studies
are needed to further evaluate these effects.
  
\section{Other Systematics}
\label{sec:other_systematics}

\subsection{Beam Chromaticity and Spectral Smoothness}
\label{sec:ant_beam_chromaticity}
By definition, the sky-averaged antenna temperature $T_{\rm ant}(\nu)$
in Equation \eqref{eq:T_ant_def} is a beam-weighted value of the sky
brightness temperature $T_{\rm sky}(\tpn)$.  Consequently, despite the
intrinsic spectral smoothness of the sky-averaged foreground spectrum,
the observed sky spectrum is corrupted by the spectral variations in
the beam patterns. This can potentially introduce unknown absorption
or emission features to the residual spectrum after removing the
foreground component using a foreground fitting model like Equation
\eqref{eq:log_poly}.

Although antenna simulations from CEM software, like CST,
HFSS\footnote{\texttt{https://www.ansys.com/Products/Electronics/ANSYS-HFSS}},
FEKO\footnote{\texttt{https://www.feko.info/}}, can provide detailed
antenna beam patterns to aid the beam chromaticity assessment and
calibration \citep[e.g., ][]{bernardi2015foreground,
  mozden2016limits}. Others have adopted a mitigation approach by
optimizing the antenna design to achieve smoother frequency response
(both in the beam pattern and antenna reflection coefficient
$\Gamma_{\rm ant}$) over a large frequency range, such as the blade
antenna from EDGES II \citep[50-100 MHz, ][]{mozden2016limits} or the
spherical monopole from SARAS 2 \citep[110-200 MHz,
][]{singh2017first}. Nonetheless, decoupling the beam dependence from
the beam-weighted measurement of $T_{\rm ant}(\nu)$ is an inverse
problem. This process is nonlinear mainly due to the lack of detailed
spatial information of the sky and the beam patterns in a single
total-power measurement.

\begin{figure}
\centering 
\includegraphics[width=.9\columnwidth]{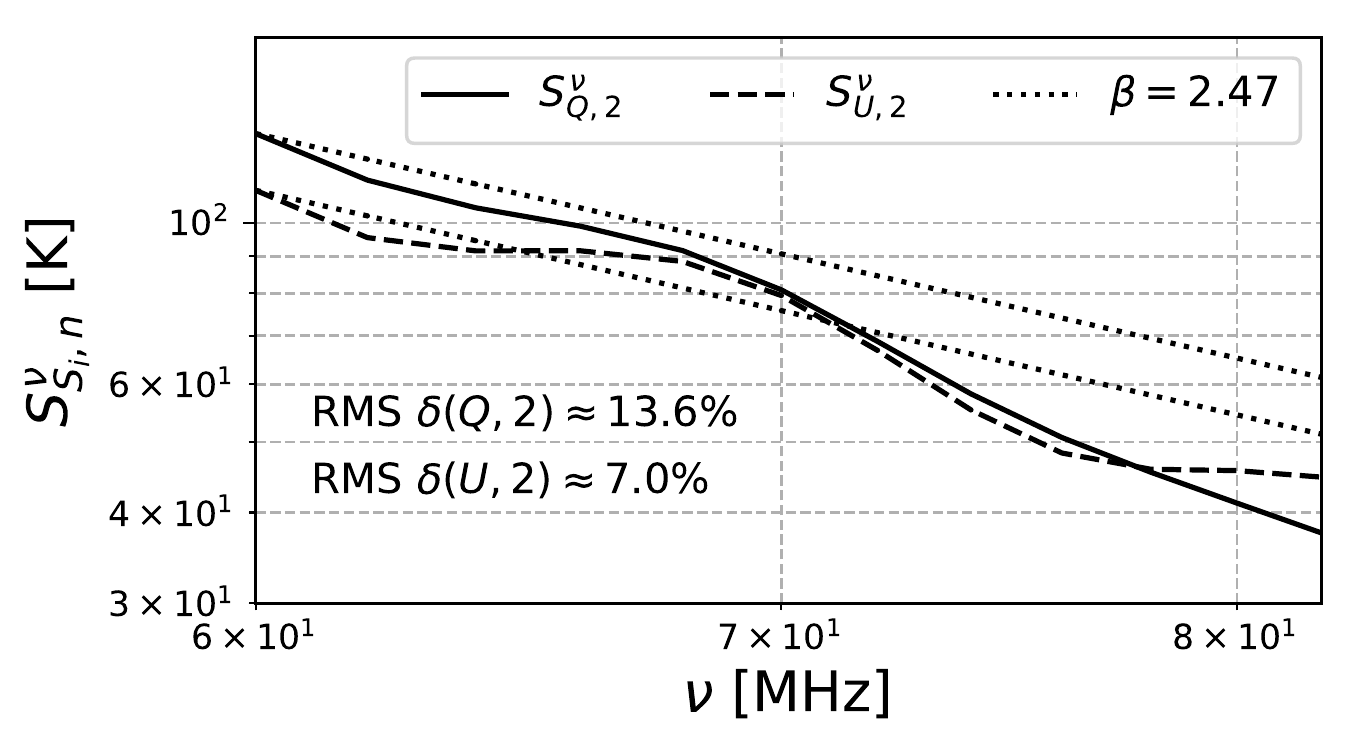}
\caption{Stokes spectra from the ground-parallel fiducial beam model
  indicate that the beam chromaticity of the antenna beam has
  distorted the smooth power-law input foreground model with spectral
  index of $\beta = 2.47$. The beam spectral structures on the Stokes
  spectra, with RMS errors between 13\% ($S_{Q,2}^{\nu}$) and 7\%
  ($S_{U,2}^{\nu}$), complicate the direct use of the Stokes spectra
  to constrain the foreground spectrum, unless the beam structures are
  understood and corrected.
}\label{fig:beam_chrom_rms_cst_beam_fiducial}
\end{figure}

Similarly, the PIPE Stokes measurements are also susceptible to
spectral variations in the antenna beams, as evident in Equations
\eqref{eq:global_stokes_I}-\eqref{eq:global_stokes_V}. In
\autoref{fig:beam_chrom_rms_cst_beam_fiducial}, PIPE simulations with
the fiducial beam indicates that the second-harmonic Stokes spectra,
$S_{Q,2}^{\nu}$ and $S_{U,2}^{\nu}$, can no longer simply track the
foreground power-law model with spectral index $\beta = 2.47$ as with
the idealized Gaussian beams.

The spectral distortions imprinted by the fiducial beam patterns on
the simulated Stokes spectra are quantified by computing an overall
RMS value across the band for the relative error between the simulated
foreground spectrum with the power law as,
\begin{equation}
  \delta(S_i,n,\nu) = 100\% \times \left|\frac{S_{S_i,n}^{\nu} - \widehat{S}_{S_i,n}^{\nu}}{\widehat{S}_{S_i,n}^{\nu}}\right|,
  \label{eq:beam_chromaticity_fom}
\end{equation}
where $\widehat{S}_{S_i,n}^{\nu_i}$ is the power law spectrum,
computed with $\beta = 2.47$,
\begin{equation}
  \widehat{S}_{S_i,n}^{\nu} = S_{S_i,n}^{\nu_0}\left(\frac{\nu}{\nu_0}\right)^{-\beta},
  \label{eq:stokes_power_law}
\end{equation}
normalized at frequency $\nu_0$, which was chosen to be the lower end
of the band, at 60 MHz. This metric simply describes the deviation
from the expected sky power law in the presence of beam chromaticity.

The RMS errors of the second-harmonic Stokes spectra relative to the
smooth power law are $13.6\%$ and $7.0\%$ for Stokes $Q$ and $U$
respectively, with the spectral gradient for this beam satisfying
$\left| \partial_{\nu}F(\tpn)\right| \ge 0.05$. These distortions,
which are contributed by the Mueller components $M_{21}(\omeganu)$ and
$M_{31}(\omeganu)$, imply that an antenna with smoother response is
needed to replace the existing sleeved dipole. Nevertheless, some
degree of beam calibration is still needed. Since correcting for the
$M_{21}$ and $M_{31}$ is also a nonlinear process, it is difficult to
account for the beam effects in the twice-diurnal spectra
analytically. A SVD-based analysis approach, as summarized in
\autoref{sec:21cm_extraction_implications}, has shown promising
potential in constraining the foreground spectrum more reliably when
simultaneously taking into account all four Stokes parameters along
with a priori training sets for foreground maps and beam models.

\subsection{Foreground Spectral Index Variations}
\label{sec:fg_spectral_index_variations}
\begin{figure}[htb!]
\centering \includegraphics[scale=.6]{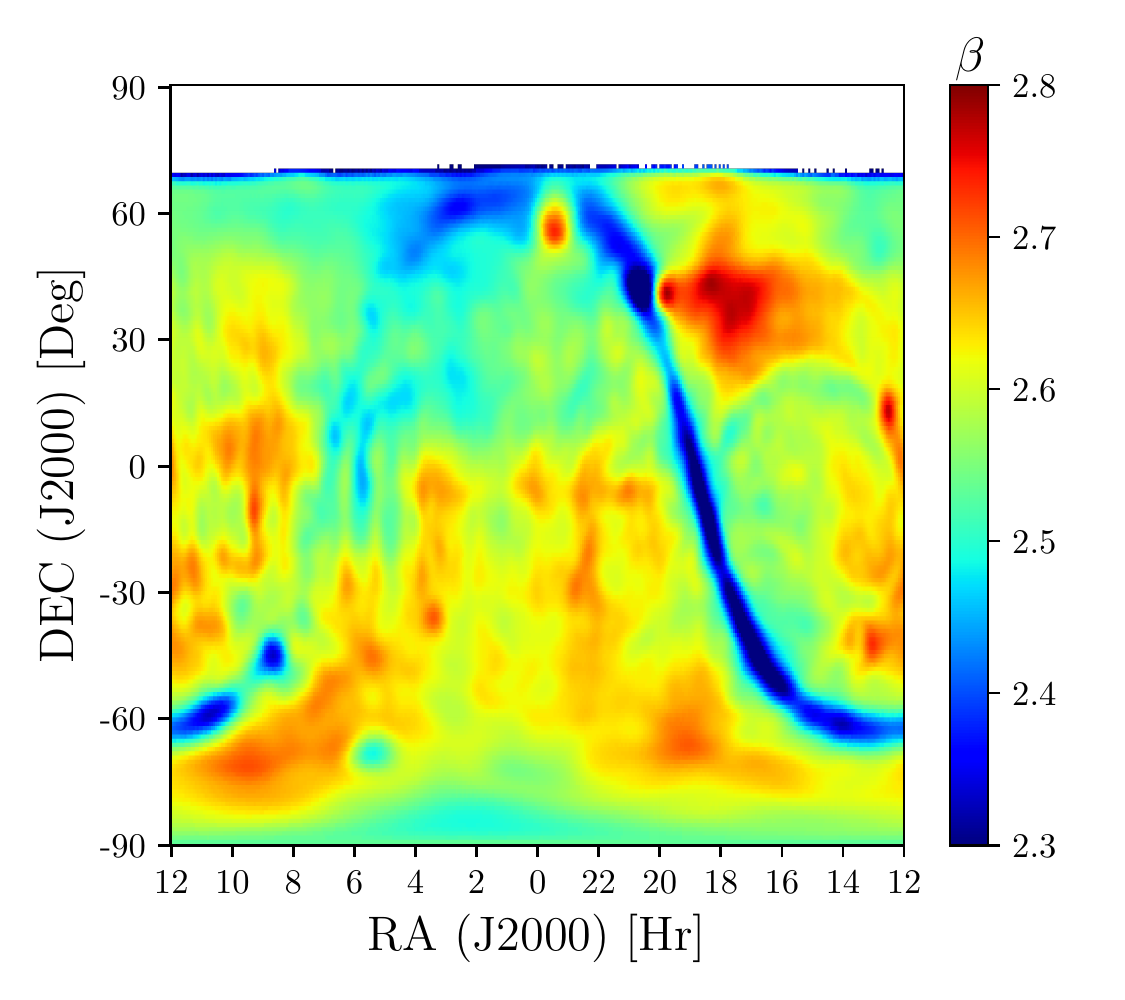}
\caption{Foreground spectral index distribution is obtained from
  extrapolating between the Haslam all-sky survey at 408 MHz and a 45
  MHz map from \cite{guzman2011all}. The missing data around the NCP
  represents $\sim 4\%$ of the whole sky in the 45-MHz
  map.}\label{fig:fg_spectral_index_distr}
\end{figure}

It is well known that the spectral index for the estimated foreground
power-law spectrum varies with position and increases with frequency,
i.e., $\beta = \beta(\omeganu)$ \citep{kogut2012synchrotron}. For
example, by extrapolating a power law between the Haslam sky map at
408 MHz to the Guzm{\'a}n map at 45 MHz \citep{guzman2011all}, it is
clear that the derived spectral indices vary between 2.3 on the
galactic plane and 2.8 at high galactic latitudes
(\autoref{fig:fg_spectral_index_distr}). By comparing different
surveys between 22 MHz and 1.4 GHz, \cite{kogut2012synchrotron} finds
that the mean spectral index increases by $\Delta\beta \sim 0.07$ per
octave in frequency. Hence, the spectrum's spectral index was assumed
constant over a small subband (such as 30-MHz band on the CTP) in the
PIPE simulation. However, $\beta$'s spatial variations are still need
to be accounted for.

It is apparent, from Equations
\eqref{eq:global_stokes_I}-\eqref{eq:global_stokes_V}, that the level
of sky modulation on the PIPE depends on the sky pointing. When the
CTP's zenith-pointing coincides with the NCP, as in the fiducial model
from \autoref{sec:fiducial_model}, only a single value of the mean
spectral index is measured in Stokes $I_{\rm net}(t_{\rm LST}, \nu)$
due to the constant FOV in this configuration. Meanwhile the remaining
three Stokes parameters are modulated by the two-fold symmetry pattern
in $M_{21}(\omeganu)$ and $M_{31}(\omeganu)$ and four-fold ones in
$M_{41}(\omeganu)$
(\autoref{fig:mueller_2x2_col1_plot_ctp1_skirt0_soil0_grid}), the
spatially dependent spectral index can affect them as
 \begin{equation}
   S_{i,\rm{net}}(\nu) \propto \int_{\Omega}\diff\Omega M_{(i+1)1}(\omeganu)
   I_{\rm sky}(\Omega,\nu_{\rm map})\left(\frac{\nu}{\nu_{\rm map}}\right)^{-\beta(\Omega)}
   \label{eq:spatially_dependent_spectral_index}
 \end{equation}
 where $S_{i,{\rm net}}$ with $i=\{1,2,3\}$ represent $\{Q_{\rm
   net},\ U_{\rm net},\ V_{\rm net}\}$, and $\nu_{\rm map}$ is the
 reference frequency at which the sky map $I_{\rm sky}$ is measured.
  
Although extrapolation between sky maps at different frequencies can
help to constrain $\beta(\omeganu)$, corrections for observational
systematics and spatial resolutions among different sky surveys can be
challenging, not to mention that not all surveys are able to cover the
entire sky like the Haslam map \citep{deoliveira2008model}. It will be
invaluable to complement them with physically motivated all-sky
models, such as the Global Sky Model \citep[GSM,
][]{deoliveira2008model}, and the Global Model for the Radio Sky
Spectrum \citep[GMOSS, ][]{rao2017gmoss}, into the analysis for the
PIPE.

\subsection{Foreground Intrinsic Polarization}
\label{sec:foreground_intrinsic_polarization}
In \autoref{sec:induced_polarization}, the PIPE is formulated by
assuming unpolarized incoming signal. The diffuse Galactic synchrotron
emission is known to be linearly polarized. However, this intrinsic
foreground polarization has been determined to be relatively weak at
arcmin-scaled angular resolution, hence is not expected to be a major
contaminant \citep{bernardi2009foregroundI,
  bernardi2010foregroundII}. Additionally, due to the ionospheric
fluctuations, the contribution from intrinsic foreground polarization
is expected to be randomized and averaged in a sky-averaged
measurement. Nonetheless, the effect of intrinsic polarization can be
characterized in future PIPE simulations simply by including the small
regional sky polarization surveys \citep[e.g.,
][]{wolleben2006absolutely, testori2008fully} or physically-motivated
all-sky polarization simulations \citep[e.g.,
][]{waelkens2009simulating, jelic2010realistic} along with the
complete $4\times4$ Mueller matrix in Equation
\eqref{eq:appdx_mueller_matrix_expand}.

\subsection{Ionospheric Effects}
\label{sec:ionospheric_effects}
Ionospheric activities are generally related to the diurnal solar
  cycle. Their bulk effects, due to electron density fluctuations, are
  enhanced during daytime, meanwhile small-scale structures are
  typically strongest during twilight and nighttime. The ionosphere
  can be considered in terms of equatorial, mid-latitudes, and polar
  regions \citep{sukumar1987ionospheric}. Ionospheric effects due to
  large-scale fluctuations in electron density are strongest in the
  equatorial and mid-latitudes, whereas effects from small-scale
  structures tend to appear more in the equatorial and polar
  regions.

There are four common ionospheric effects that need to be considered:
refraction, absorption, attenuation, and scintillation. Total electron
content (TEC) fluctuations in the large scale structure of the
ionosphere ($\sim$ 10-100s km) can alter transmission properties of
the medium and propagation direction of the incoming radio
signal. Some of the previous studies have suggested that ionospheric
refraction and absorption due to TEC fluctuations can have long-term
effects on the sensitivity of the observations, such as increase in
$1/f$ noise \citep{vedantham2013chromatic,
  datta2016effects}. Meanwhile, another study has pointed out that
such variability does not pose a significant problem for ground-based
global 21 cm experiments as long as the data are randomized and
averaged at a time interval shorter than the characteristic time scale
of the ionospheric $1/f$ noise
\citep{sokolowski2015impact}. Nonetheless, since the harmonic analysis
of the PIPE Stokes relies on concatenating multiple consecutive days
of data, long-term effects of the ionospheric fluctuations will need
to be characterized and corrected. One possible remedy is to align and
randomize the data of each day in LST. The sky-modulated PIPE
waveforms should be preserved.

\begin{figure}
\centering 
\includegraphics[width=0.9\columnwidth]{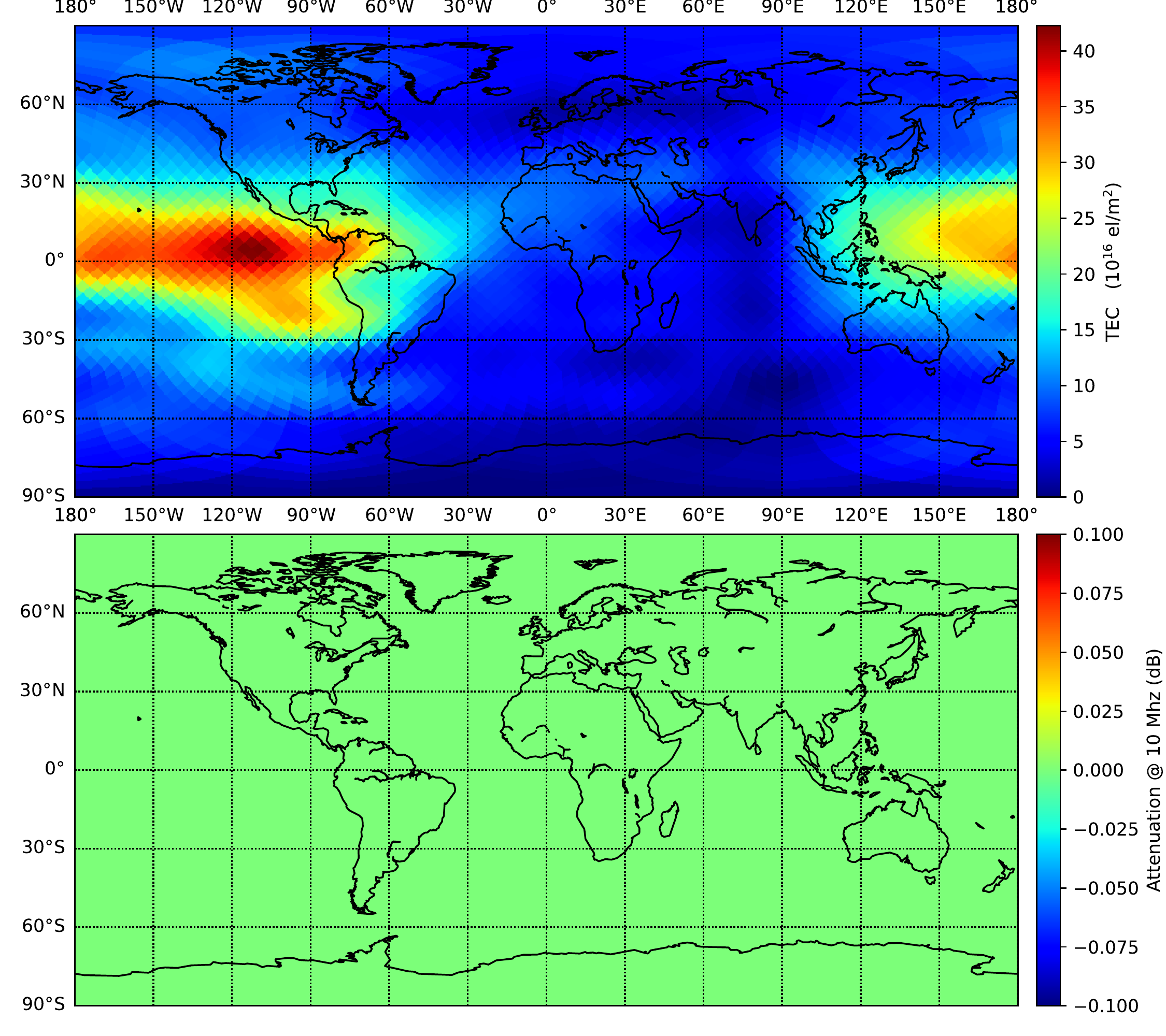}
\caption{TEC (top) and attenuation (bottom) maps for the ionosphere
  conditions on normal day (Sept 01, 2017 at 00:00 UTC). Although
  there are large variations TEC close to the Equator, which is
  typical, no significant attenuation is observed. Source: CODE and
  NOAA.}\label{fig:tec_atten_20170901_0000}
\end{figure}

\begin{figure}
\centering 
\includegraphics[width=0.9\columnwidth]{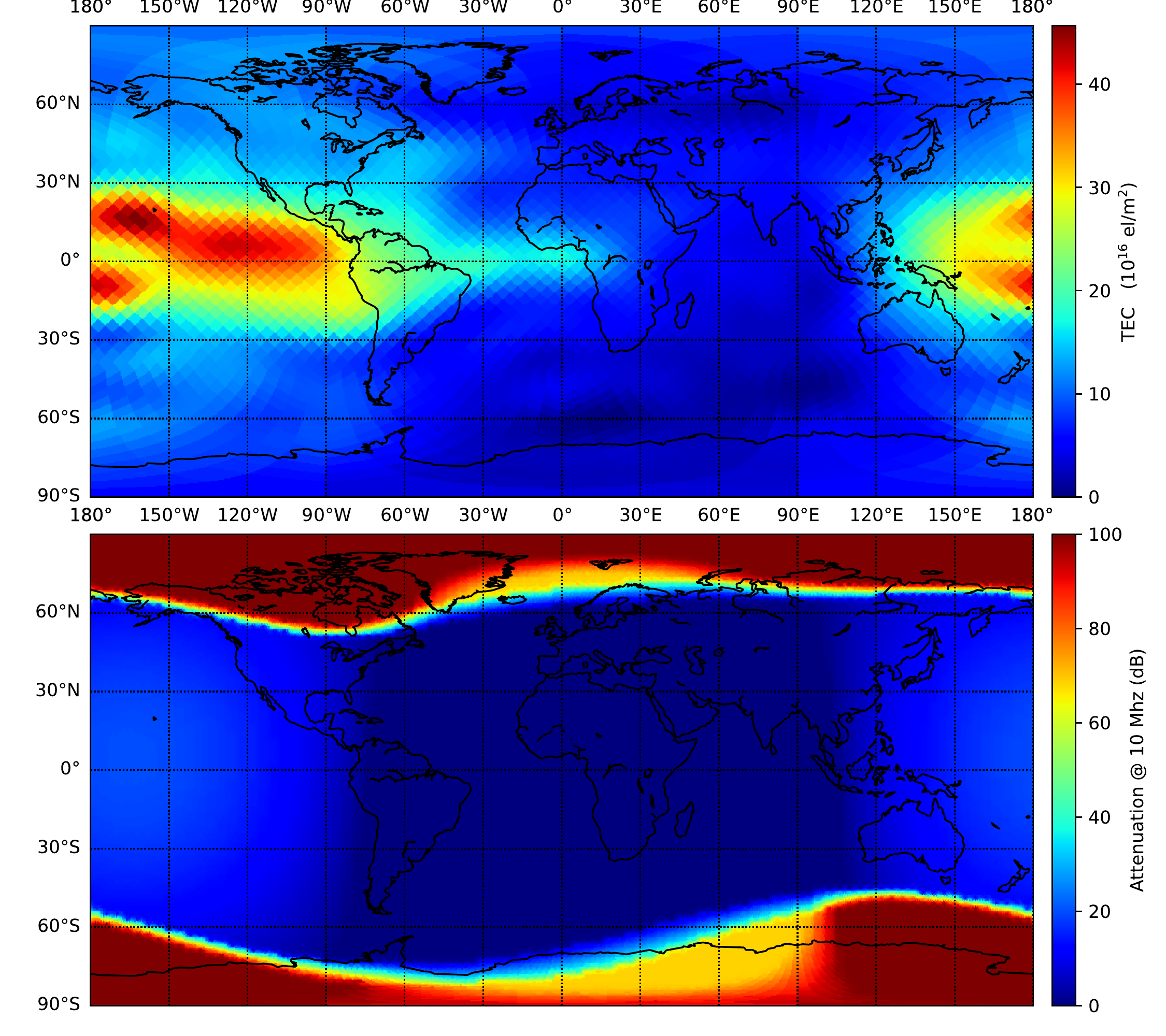}
\caption{TEC (top) and attenuation (bottom) maps for the ionosphere
  conditions on the day with a strong solar flare (Sept 09, 2017 at
  23:00 UTC). During the active period, ionosphere in both polar
  regions (latitude $\geq 60^{\circ}$) are saturated with high energy
  particles so that the signal are large attenuated. Although this is
  not ideal for any experiments located at those latitudes, these
  solar activities are neither frequent nor long-lasting. Hence, it is
  not a major threat for any ground-based experiments close to the
  poles. Source: CODE and NOAA.}\label{fig:tec_atten_20170909_2300}
\end{figure}

On the other hand, if one desires to mitigate the horizon obstruction
by observing at latitudes closer to the geographic poles, attenuation
from the ionosphere will be more dominant, especially during strong
solar activity such as flares and coronal mass ejections (CME). By
comparing the archival TEC data\footnote{Data provided by the Center
  for Orbit Determination in Europe (CODE), fetched by the Python
  script \texttt{radionopy} provided by Prof. James Aguirre from the
  University of
  Pennsylvania,\\ \texttt{https://github.com/UPennEoR/radionopy}}
along with the attenuation maps\footnote{Attenuation maps are based on
  \cite{sauer2008global} data, acquired from the National Oceanic and
  Atmospheric Administration
  (NOAA):\\ \texttt{https://www.swpc.noaa.gov/content/global-d-region-\\absorption-prediction-documentation}}
between a normal day (Sept 01, 2017 at 00:00 UTC,
\autoref{fig:tec_atten_20170901_0000}) and a day with a strong flare
(Sept 09, 2017 at 23:00 UTC, \autoref{fig:tec_atten_20170909_2300}),
it is apparent that majority of the ionosphere in polar regions is
saturated during the flare day, which leads to complete attenuation of
the sky signal. However, the occurrence of these events is infrequent,
hence can be flagged and removed from the data if needed. Note that
the attenuation and TEC level do not necessarily correlate.

Furthermore, there is also ionospheric scintillation, which
  appears as small scale variations ($\sim$1 m to tens of km), can cause
  rapid disruptions on the RF signal's amplitude and phase coherency
  \citep{vanbemmel2007ionospheric} and result in noise-like terms in
  the observed spectrum. These variations may also help to scramble
  the intrinsic foreground polarization through changes in the amount
  of Faraday rotation as mentioned in the last section.

\section{Implications on 21 cm Background Signal Extraction}
\label{sec:21cm_extraction_implications}
By now, it is evident that the PIPE is more complex than the
simplistic Gaussian beam simulation in NB17, especially in the
presence of instrumental systematics on the beams and intrinsic
spectral properties of the foreground synchrotron emission. Other
global experiments have utilized elaborate receiver calibration
schemes, detailed CEM beam models, and antenna designs with smoother
responses, to mitigate spectral distortions on the observed
spectrum. Nonetheless, uniqueness can be a concern when trying to
correct for all these systematics from a single averaged total-power
spectrum.

One advantage of the sky-modulated PIPE over a single total-power
spectrum is the additional constraints on the foreground and antenna
beam characteristics imprinted in the full-Stokes measurement. In
principle, if all four Stokes measurements are constrained
simultaneously, the robustness and accuracy of separating of the
foreground spectrum from the weak 21 cm signal will
improve. Furthermore, underlying instrumental and observational
systematics can be accounted for if the observed signal can be
decomposed into different components (or modes) and compared to a
priori information of the expected systematics. An SVD-based analysis
provides such a means.

In brief, the SVD algorithm decomposes the observed sky-modulated PIPE
signal and simulation into two different sets of eigenmodes. Ones
corresponding to the foreground and systematics in the observation can
be isolated and removed based on the modes appear in the simulated
data. Statistical uncertainties of the systematic components
introduced to the simulated data are constrained by a priori training
data sets. Subsequently, remaining eignmodes in the observed signal
allow the background signal of interest to be reconstructed and
statistically bounded by theoretical global 21 cm models.

\begin{figure}[!htb]
\includegraphics[width=\columnwidth]{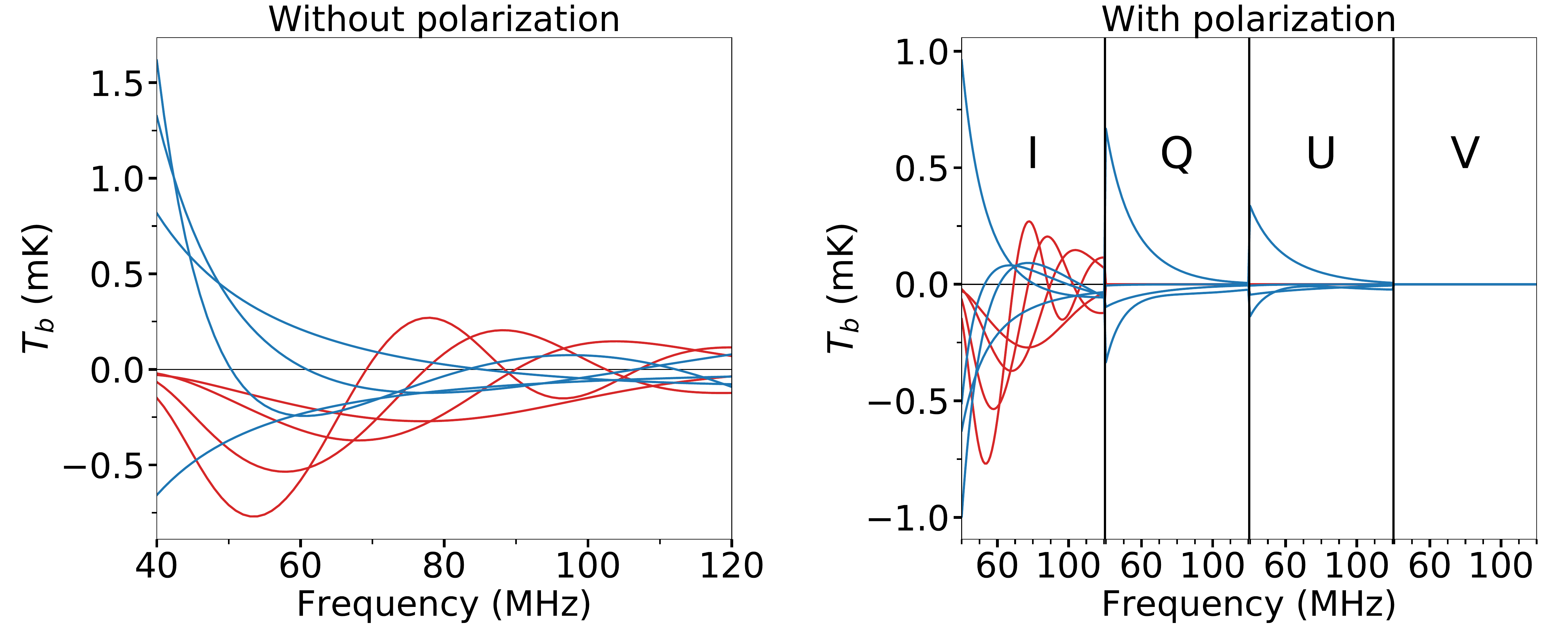}
\caption{Illustration of the SVD basis functions for the simulated
  total power alone (top), and with induced polarization
  (bottom). When decomposing the total power alone, the overlapping
  between the modes for injected 21 cm background signal (red) and the
  foreground spectrum (blue). However, since the foreground is solely
  responsible for the induced polarization, the SVD modes in Stokes
  $Q$ and $U$ provide the needed additional constraints on the
  foreground.}\label{fig:svd_mode_comparison_plot}
\end{figure}

As an example, the SVD algorithm, as implemented in the Python code
\texttt{pylinex}\footnote{\texttt{https://bitbucket.org/ktausch/pylinex}}
\citep{tauscher2018global}, was applied to the fiducial PIPE
simulation described in \autoref{sec:fiducial_model}, embedded with an
additional global 21 cm model generated by the
\texttt{ARES}\footnote{The Accelerated Reionization Era Simulation,
  \texttt{https://bitbucket.org/mirocha/ares}}
\citep{mirocha2014decoding} code. Statistical training sets for the
beam-weighted foreground were constructed by convoluting the scaled
Haslam map with a series of CST beams consisting of different model
perturbations (e.g., thermal expansions on different components on
antenna structure, ground soil properties and thickness, ground screen
characteristics). Meanwhile, a collection of global 21 cm models was
generated using the ARES code with different combinations of neutral
hydrogen fraction $x_{\rm HI}(z)$ and 21 cm spin temperatures $T_{s,
  {\rm 21cm}}(z)$. Also, as a blind test, the specific global 21 cm
model embedded in the PIPE test signal was left out of the training
set.

\begin{figure}[!htb]
\centering 
\includegraphics[width=\columnwidth]{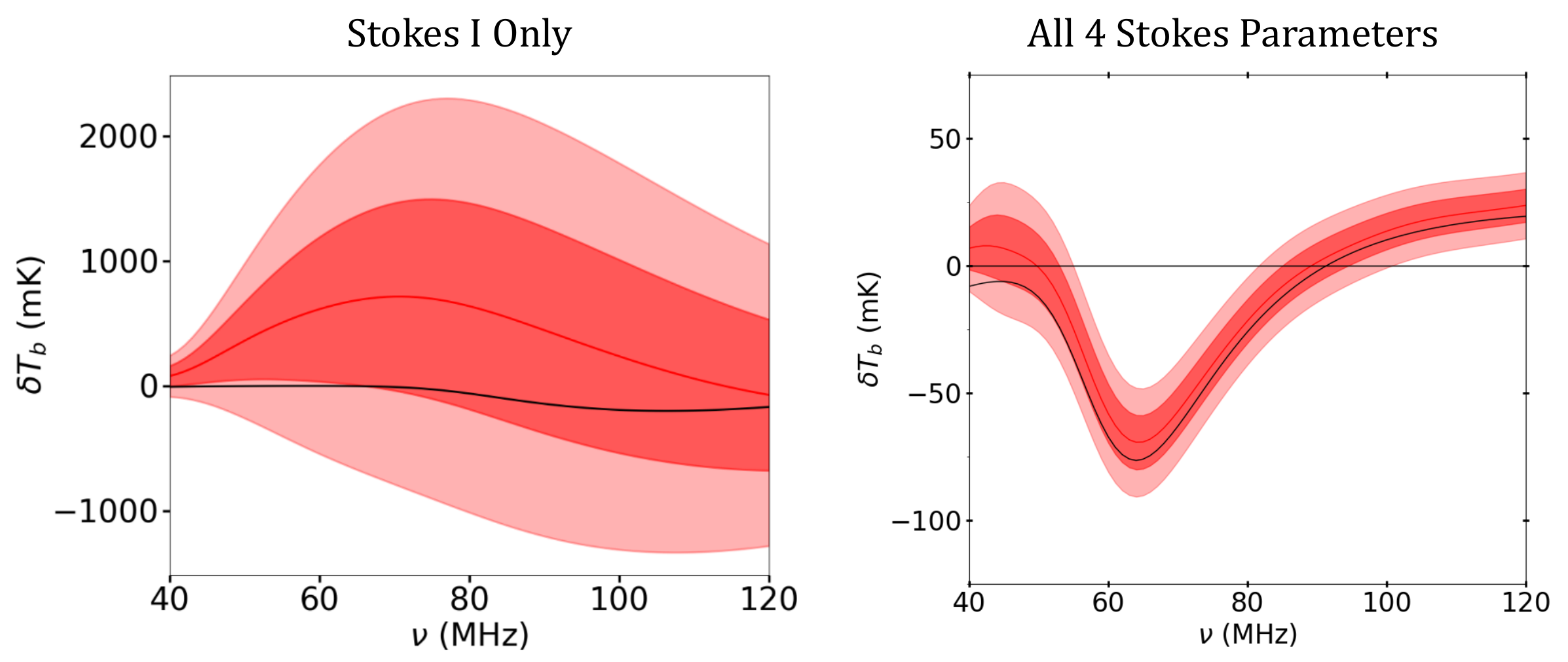}
\caption{By utilizing a large training set of global 21 cm signal,
  simulated beam models, and the Haslam map, the SVD-based
  \texttt{pylinex} code demonstrates the power of the PIPE approach
  when simultaneously constraining all four Stokes parameters (right)
  instead of just the total-power spectrum as in the conventional
  approach (left). The two uncertainty bands are $1\sigma$ (dark red)
  and $2\sigma$ (light red) of the recovered signal (solid red curve),
  overplotted on top of the input global 21 cm model (solid black
  curve). Note also the scales between the two panels are at least an
  order of magnitude different.}\label{fig:svd_extraction_compare}
\end{figure}

The SVD eigenmodes of a simulated single total-power spectrum are
contrasted with the full-Stokes counterpart in
\autoref{fig:svd_mode_comparison_plot}. When the additional Stokes
data are included, distinctions between the foreground and background
signal's eigenmodes are more prominent as the eignmodes corresponding
to the foreground components only present in Stokes $Q$ and
$U$. Hence, the foreground and beam effects can be identified more
easily. As a result, the reconstructed 21 cm signal from the
all-Stokes spectra is more robust and precise than the single
total-power spectrum, as illustrated in
\autoref{fig:svd_extraction_compare}.

The robustness of this analysis for separating unwanted data
components from the 21 cm background can be quantified by checking the
goodness of the fit to the data through either the traditional reduced
chi-squared statistic or the newly devised psi-squared statistic
\citep{tauscher2018new}. If a goodness of fit threshold is not met,
modifications to the PIPE simulation, such as including extra
systematics and training sets, may be required before the SVD analysis
process restarts. Additionally, determining an optimal number of modes
for each of the signal components plays a key role in the success of
the SVD approach. \cite{tauscher2018global} show that minimizing the
Deviance Information Criterion (DIC) best optimizes the number of
modes to use for each training set or data component. Follow-up
studies are under way to apply similar SVD analysis on new observation
data.

\section{Conclusion and Future Work}
\label{sec:conclusion_future_work}
In this study, we review different implementation aspects of the newly
proposed polarimetry-based observational technique, using sky
modulation for the Projection-Induced Polarization Effect (PIPE), to
constrain the foreground spectrum for global 21 cm
cosmology. Simulation of the PIPE has been extended to include
realistic beam effects, such as beam chromaticity, beam distortions
from ground soil, and Earth's horizon obstruction. In conjunction, the
Cosmic Twilight Polarimeter (CTP) is presented as a testbed instrument
for implementing the network-theory based calibration scheme and the
sky-modulated PIPE observation by tilting a broadband sleeved dipole's
pointing toward the NCP at a latitude of $38^{\circ}$N.

Instead of reference-load switching, the CTP has adopted a
network-theory based calibration scheme to correct for the power
transducer gain, $G_T(\nu)$ and noise temperature, $T_n(\nu)$.
Instead of attempting to determine these two variables directly, since
both are functions of the antenna reflection coefficient $\Gamma_{\rm
  ant}(\nu)$, two sets of intrinsic system variables ($S$-parameters
and noise parameters) are measured. Subsequently, $G_T(\nu)$ and
$T_n(\nu)$ of the observing system are computed by substituting
$\Gamma_{\rm ant}(\nu)$ into the parametrization formulae of the
network parameters.

Despite strong RFI contamination across the 60-90 MHz band and most of
the observational data, by duplicating and concatenating one of the
cleanest days of data to emulate a continuous observational data set,
a desired FFT resolution for the low-order harmonics was
achieved. Preliminary analysis of the harmonics, compared to the PIPE
simulation, suggests the presence of a twice-diurnal component in the
net Stokes parameters around 82 MHz, specifically in the Stokes
$U$. Based on the estimated statistics in the observation, S/N of the
twice-diurnal components in Stokes $Q$ and $U$ are 3.27 and 7.40
respectively. Importantly, the observed harmonics in Stokes $U$ are
consistent with the PIPE simulation since their ratios for $n =
\{1,\ 2,\ 3\}$ are unity within uncertainty. However, there are
discrepancies between the observation and simulation, such as ones in
other Stokes parameters $Q$ and $V$, mainly due to RFI corruptions,
small sample size, and unknown systematics.

Among the discussed systematics, the following contribute the most to
spectral distortions in the sky-modulated PIPE measurements:
\begin{enumerate}     
\item Similarly to the total-power measurement, the beam chromaticity
  also affects the PIPE Stokes antenna temperatures. Although
  dynamical sky modulation provides a unique way to constrain the
  foreground component, beam chromaticity imprinted in the Stokes
  parameters needs to be corrected. Hence, constraining the foreground
  spectrum using the Stokes parameters becomes more complicated than
  the idealized procedure proposed in NB17.
  
\item Simply tilting the antenna forward, as in the CTP prototype
  presented here, is not recommended when the antenna is close to the
  ground. Nearfield interactions with the ground soil can corrupt the
  beam pattern's smoothness (\autoref{sec:ground_distortions}). Having
  the antenna situated on a slope or suspended off the ground when
  tilted will alleviate these corruptions. Additionally, FOV
  obstruction from the Earth's horizon can be reduced by relocating
  the instrument closer to the geographic poles, thus decreasing the
  antenna's tilting angle and suppressing the ground interactions.

\item Spatial distribution of the foreground spectral index
  $\beta(\Omega)$ also affects the accuracy of the recovered
  foreground spectrum using the sky-modulated PIPE. When no longer
  sharing a uniform spectral index across the sky, each Stokes
  parameter is only sensitive to the given sky region modulated by the
  corresponding Mueller pattern, as described in Equation
  \eqref{eq:spatially_dependent_spectral_index}. Hence, the net Stokes
  $Q$ and $U$ can produce foreground spectra with slightly different
  frequency dependence than the foreground in Stokes $I$.
\end{enumerate}

Nonetheless, the SVD-based \texttt{pylinex} code provides a promising
background signal extraction procedure when using statistical training
sets to address some of the intrinsic foreground and instrumental
systematics, such as the three above. Preliminary analysis on the
simulated data indicates that it is plausible to extract the
background 21 cm signal as long as training sets with a priori
information on the foreground, antenna beam, electronics calibration,
and other systematics are sufficiently well defined (through sky
survey maps and models like GMOSS, CST beam models, and laboratory
measurements of $S-$ and noise-parameters, respectively).

Instead of requiring precise knowledge of the foreground and
instrumental effects, \texttt{pylinex} constrains the background 21 cm
signal by decomposing the training sets of each possible data
component into eigenmodes. By simultaneously fitting them to the
full-Stokes measurements, the algorithm provides a well-informed
separation between background, foreground, and systematics. The
goodness of fit to the observation can be optimized with the reduced
chi-squared or psi-squared statistics. The order of SVD eigenmodes to
be used for each training set is optimally determined by minimizing an
information criterion such as the DIC.

Although preliminary CTP observation has shown evidence of
sky-modulated PIPE, a follow-up effort is necessary to improve the
sensitivity of the data, such as relocating the system to the Green
Bank Observatory (GBO, $38.4^{\circ}$N, $79.8^{\circ}$W) inside the
radio quiet zone in West Virginia. Although the presented calibration
is sufficient for the induced polarization, the CTP system is
currently being upgraded and refined to improve the sensitivity of the
gain and noise calibration for the global 21 cm science (which is at
least three orders of magnitude lower than the induced
polarization). Detailed component-level network models are being
constructed for the frontend and receiver using the Keysight's
Advanced Design Simulation (ADS) electronic design software. The
network models will help to validate the calculated transducer gain
and noise temperature when the input ports are subjected to the
antenna impedance. Additionally, a new broadband and dual-polarized
antenna design with a smoother frequency response is being explored to
reduce spectral distortions from the beam pattern, along with
reviewing possible modifications on the existing sleeved dipole for
comparison.

Another possible path, to mitigate the ground distortions on the
tilted beam, is to point the antenna at the zenith instead of at the
celestial pole when observing at the GBO. This approach will
significantly decrease the sensitivity in detecting the twice diurnal
waveforms in the Stokes parameters compared to when observing at the
poles. However, the pattern recognition technique implemented in
\texttt{pylinex}, in combination with suitable training sets, may be
able to take advantage of the complete information provided by
full-Stokes measurements. Although this may resemble conventional
global 21 cm experiments, to the best of our knowledge, no other
single-element total-power global experiment has incorporated
polarimetry and pattern recognition together. With the upgraded CTP,
applications of a pattern recognition method on both the
drift-scanning total-power and the dynamic PIPE measurements can be
fully evaluated, for the first time with observational data.

Moreover, there is also great potential for adopting the
induced polarization approach on a future space-based mission since
many of the ground-based challenges, such as ground interactions with
the antenna beam and horizon obstruction, as well as ionospheric
effects, will be eliminated. In fact, previous studies \citep[e.g.,
][]{burns2012probing, burns2017space, falcke2018radio} have suggested
that the lunar farside is the optimal location for such an experiment
due to its pristine radio-quiet environment within the inner solar
system.

\section{Acknowledgments}
\label{sec:acknowledgments}
The National Radio Astronomy Observatory is a facility of the National
Science Foundation operated under cooperative agreement by Associated
Universities, Inc. Support for this work was provided by the NSF
through the Grote Reber Fellowship Program administered by Associated
Universities, Inc./National Radio Astronomy Observatory. This research
was also supported by the NASA Ames Research Center via Cooperative
Agreements NNA09DB30A, NNX15AD20A, and NNX16AF59G to
J. O. Burns. D. Rapetti is supported by a NASA Postdoctoral Program
Senior Fellowship at the NASA Ames Research Center, administered by
the Universities Space Research Association under contract with
NASA. This work was also directly supported by the NASA Solar System
Exploration Virtual Institute cooperative agreement 80ARC017M0006. The
authors gratefully acknowledge the Equinox Farm, LLC. for hosting and
providing utilities for the CTP. B. Nhan would also like to thank
K. Makhija for his comments on the instrumentation description.

\appendix
\section{A. Conversion between Jones and Mueller Matrices}
\label{appdx:jones2mueller}
For a given dual polarization antenna, the antenna beam pattern can be
described in terms of a $2\times2$ Jones matrix, with the $(\tpn)$
notations suppressed for the ease of reading, as,
\begin{equation}
  \begin{aligned}
    \bm{J}_{\rm ant} &= 
    \begin{pmatrix}
      J_{11} & J_{12}\\
      J_{21} & J_{22}
    \end{pmatrix} = 
    \begin{pmatrix}
      |E_{\theta}^{X}|e^{i\Phi_{\theta}^{X}} & |E_{\phi}^{X}|e^{i\Phi_{\phi}^{X}} \\
      |E_{\theta}^{Y}|e^{i\Phi_{\theta}^{Y}} & |E_{\phi}^{Y}|e^{i\Phi_{\phi}^{Y}}
    \end{pmatrix},
  \end{aligned}
  \label{eq:appdx_jones_matrix}
\end{equation}
where $|E|$ and $\Phi$ are the magnitude and phase of the $\theta$ and
$\phi$ components of the farfield pattern. The output signal
$E$-fields received by both polarization the antenna from the incoming
$E$-fields is given by $\bm{E}_{\rm out} = \bm{J}_{\rm ant}\bm{E}_{\rm
  in}$ or in matrix form,
\begin{equation}
  \begin{pmatrix}
    E_X \\ E_Y
  \end{pmatrix} =
  \begin{pmatrix}
      J_{11} & J_{12}\\
      J_{21} & J_{22}
  \end{pmatrix}
  \begin{pmatrix}
    E_x \\ E_y
  \end{pmatrix}_{\rm in}.
  \label{eq:appdx_eout_jones_ein}
\end{equation}

Subsequently, the coherency vectors for the output and input signal
can be computed by taking the outer product of the $E$-fields and
their complex conjugates as $\bm{C} = \bm{E}\otimes\bm{E}^*$, similar
to Equation \eqref{eq:appdx_coherency_vector}, to produce the
relation,
\begin{equation}
  \bm{C}_{\rm out} = (\bm{J}_{\rm ant}\otimes\bm{J}_{\rm ant}^*)\bm{C}_{\rm in}.
  \label{eq:appdx_coherency_vector_in2out}
\end{equation}
Using the definition of the Stokes parameters as the linear
combination of the autocorrelation and cross-correlation of the
$E$-fields, the Stokes vector can be written in terms of the coherency
vector as $\bm{S} = \bm{A}\bm{C}$ for both the input and output
signal, where
\begin{equation}
  \bm{A} = 
  \begin{pmatrix}
    1 & 0 & 0 & 1 \\
    1 & 0 & 0 & -1 \\
    0 & 1 & 1 & 0 \\
    0 & i & -i & 0
  \end{pmatrix},
  \label{eq:appdx_stokes_A_matrix}
\end{equation}
to obtain the relation between the input and output Stokes vectors as,
\begin{equation}
  \bm{S}_{\rm out} = \bm{A}(\bm{J}_{\rm ant}\otimes\bm{J}_{\rm ant}^*)\bm{A}^{-1}\bm{S}_{\rm in} = \bm{M}_{\rm ant}\bm{S}_{\rm in},
  \label{eq:appdx_stokes_vector_in2out}
\end{equation}
where $\bm{M}_{\rm ant}$ is the $4\times4$ antenna Mueller
matrix. Therefore,
\begin{align}
    \bm{M}_{\rm ant} &=  \bm{A}(\bm{J}_{\rm ant}\otimes\bm{J}_{\rm ant}^*)\bm{A}^{-1}\\
    & =\begin{pmatrix}
    0.5(E_1 + E_2 + E_3 + E_4) & 0.5(E_1 - E_2 - E_3 + E_4)  & F_{13} + F_{42}  & -G_{13} - G_{42}\\
    0.5(E_1 - E_2 + E_3 - E_4) & 0.5(E_1 + E_2 - E_3 - E_4)  & F_{13} - F_{42}  & -G_{13} + G_{42}\\
    F_{14} + F_{32} & F_{14} - F_{32}  & F_{12} + F_{34} & -G_{12} + G_{34}\\
    G_{14} + G_{32} & G_{14} - G_{32}  & G_{12} + G_{34} & F_{12} - F_{34} 
    \end{pmatrix},
  \label{eq:appdx_mueller_matrix_expand}
\end{align}
where 
\begin{align}
  E_k &= J_kJ_k^*, \\
  F_{kl} &= F_{lk} = {\rm Re}(J_kJ_l^*) = {\rm Re}(J_k^*J_l), \\
  G_{kl} &= -G_{lk} = {\rm Im}(J_kJ_l^*) = -{\rm Im}(J_k^*J_l),
  \label{eq:muller_matrix_terms}
\end{align}
for $k,l \in \{1,\ 2,\ 3,\ 4\}$, and $\{J_1,\ J_2,\ J_3,\ J_4\}$
respectively representing the four components
$\{J_{11},\ J_{22},\ J_{12},\ J_{21}\}$ of the Jones matrix. A more
detailed derivation of the conversion between Jones and Mueller
matrices can commonly be found in literature, such as one provided in
the Appendix 4 of \cite{fujiwara2007spectroscopic}.

\section{B. Calibration for Stokes Antenna Temperature}
\label{appdx:stokes_calibration}
To derive the formulae to convert the correlation spectra into
equivalent temperature units, we first note that the complex
$E$-fields for each polarization in the coherency vector,
\begin{equation}
  \bm{C}(t,\nu) = 
  \begin{pmatrix}
    E_XE_X \\ E_XE_Y\\
    E_YE_X \\ E_YE_Y 
  \end{pmatrix}_{(t,\nu)},
  \label{eq:appdx_coherency_vector}
\end{equation}
are equivalent to the complex antenna voltages,
$\widetilde{V}_X(t,\nu)$ and $\widetilde{V}_Y(t,\nu)$, which are
acquired from Fourier transforming the sampled antenna
voltages. Meanwhile, the power transducer gain $G_T(t,\nu)$ is defined
as the magnitude square of the complex voltage gain $g_T(t,\nu)$,
i.e., $G_T = |g_T|^2$. The noise power $P_n(\nu,t_t)$ from the FE signal
path can be written as the absolute square of the complex noise
voltage $\widetilde{V}_n(t,\nu)$, and $P_n(t,\nu) = \langle
|\widetilde{V}_n\widetilde{V}_n^*|\rangle = k_B\Delta\nu T_n(t,\nu)$
with the Boltzmann constant $k_B$. Hence the monochromatic $E$-fields
of both polarizations can be parametrized as,
\begin{align}
  E_X(t,\nu) = \widetilde{V}_X(t,\nu) &=
  g_{T,X}(t,\nu)\left[\widetilde{V}_{{\rm ant},X}(t,\nu) +
  \widetilde{V}_{n,X}(t,\nu) \right],\\
  E_Y(t,\nu) = \widetilde{V}_Y(t,\nu) &=
  g_{T,Y}(t,\nu)\left[\widetilde{V}_{{\rm ant},Y}(t,\nu) +
  \widetilde{V}_{n,Y}(t,\nu) \right],
  \label{eq:appdx_voltage_signalXY}
\end{align}
where $\widetilde{V}_{{\rm ant}}(t,\nu)$ is the antenna voltage
signal of each polarization when observing the sky, where $P_{\rm
  ant}(t,\nu) = \langle|\widetilde{V}_{\rm ant}\widetilde{V}_{\rm
  ant}^*|\rangle = k_B\Delta\nu T_{\rm ant}(t,\nu)$. With these two
equations, we can write the correlation terms in the coherency vector
as,
\begin{align}
  \langle \widetilde{V}_X\widetilde{V}_X^* \rangle &= G_{T,X}\left(\langle \widetilde{V}_{{\rm
      ant},X}\widetilde{V}_{{\rm ant},X}^*\rangle + \langle \widetilde{V}_{n,X}\widetilde{V}_{n,X}^*\rangle
  + \langle \widetilde{V}_{{\rm ant},X}\widetilde{V}_{n,X}^*\rangle + \langle \widetilde{V}_{{\rm
      ant},X}^*\widetilde{V}_{n,X}\rangle \right),\\
  \langle \widetilde{V}_Y\widetilde{V}_Y^* \rangle &= G_{T,Y}\left(\langle \widetilde{V}_{{\rm
      ant},Y}\widetilde{V}_{{\rm ant},Y}^*\rangle + \langle \widetilde{V}_{n,Y}\widetilde{V}_{n,Y}^*\rangle
  + \langle \widetilde{V}_{{\rm ant},Y}\widetilde{V}_{n,Y}^*\rangle + \langle \widetilde{V}_{{\rm
      ant},Y}^*\widetilde{V}_{n,Y}\rangle\right),\\
  \langle \widetilde{V}_X\widetilde{V}_Y^* \rangle &= \sqrt{G_{T,X}G_{T,Y}}\left(\langle \widetilde{V}_{{\rm
      ant},X}\widetilde{V}_{{\rm ant},Y}^*\rangle + \langle \widetilde{V}_{n,X}\widetilde{V}_{n,Y}^*\rangle
  + \langle \widetilde{V}_{{\rm ant},X}\widetilde{V}_{n,Y}^*\rangle + \langle \widetilde{V}_{{\rm
      ant},Y}^*\widetilde{V}_{n,X}\rangle\right),\\
  \langle \widetilde{V}_Y\widetilde{V}_X^* \rangle &= \sqrt{G_{T,X}G_{T,Y}}\left(\langle \widetilde{V}_{{\rm
      ant},X}\widetilde{V}_{{\rm ant},Y}^*\rangle + \langle \widetilde{V}_{n,Y}\widetilde{V}_{n,X}^*\rangle
  + \langle \widetilde{V}_{{\rm ant},Y}\widetilde{V}_{n,X}^*\rangle + \langle \widetilde{V}_{{\rm
      ant},X}^*\widetilde{V}_{n,Y}\rangle\right),
  \label{eq:appdx_volt_corr}
\end{align}
where the $(t,\nu)$ notation has been suppressed for the ease of
reading, and $\langle\ldots\rangle$ represents an ensemble
average. The equivalent antenna temperatures resulting from these
auto- and cross-powers can be written as
\begin{equation}
  T_{{\rm ant},kl}(t,\nu) = \frac{\langle \widetilde{V}_{{\rm ant},k}\widetilde{V}_{{\rm
        ant},l}^*\rangle}{k_B\Delta\nu},
\label{eq:appdx_ant_temp_corr}
\end{equation}
where the subscripts $\{k, l\}$ correspond to the $X$ or $Y$ polarizations. 

Subsequently, the calibrated Stokes parameters, which were computed
with the auto-spectra and cross-spectra, can be written in temperature
unit as the following,
\begin{flalign}
  I_{\rm cal}(t,\nu) ={} &T_{{\rm ant},XX}(t,\nu) + T_{{\rm ant},YY}(t,\nu) \nonumber\\ ={}
  &\frac{1}{k_B\Delta\nu}\left[ \left(\frac{\langle
      \widetilde{V}_X\widetilde{V}_X^*\rangle}{G_{T,X}} +
    \frac{\langle
      \widetilde{V}_Y\widetilde{V}_Y^*\rangle}{G_{T,Y}}\right) -
    \left(T_{n,X}+T_{n,Y}\right) - 2{\rm Re}\left(\langle
    \widetilde{V}_{{\rm ant},X}\widetilde{V}_{n,X}^*\rangle \right) -
    2{\rm Re}\left(\langle \widetilde{V}_{{\rm
        ant},Y}\widetilde{V}_{n,Y}^*\rangle\right) \right], \\  
  Q_{\rm cal}(t,\nu) ={} &T_{{\rm ant},XX}(t,\nu) - T_{{\rm ant},YY}(t,\nu) \nonumber\\ ={}
  &\frac{1}{k_B\Delta\nu}\left[ \left(\frac{\langle
      \widetilde{V}_X\widetilde{V}_X^*\rangle}{G_{T,X}} -
    \frac{\langle
      \widetilde{V}_Y\widetilde{V}_Y^*\rangle}{G_{T,Y}}\right) -
    \left(T_{n,X}-T_{n,Y}\right)- 2{\rm Re}\left(\langle
    \widetilde{V}_{{\rm ant},X}\widetilde{V}_{n,X}^*\rangle \right) +
    2{\rm Re}\left(\langle \widetilde{V}_{{\rm
        ant},Y}\widetilde{V}_{n,Y}^*\rangle\right) \right],
   \label{eq:appdx_stokes_calIQ}
\end{flalign}
\begin{flalign}
   U_{\rm cal}(t,\nu) ={} &T_{{\rm ant},XY}(t,\nu) + T_{{\rm ant},YX}(t,\nu) \nonumber\\ ={}
  &\frac{2}{k_B\Delta\nu}\left[\frac{{\rm Re}\left(\langle
      \widetilde{V}_X\widetilde{V}_Y^*\rangle\right)}{\sqrt{G_{T,X}G_{T,Y}}}
    - {\rm Re}\left(\langle
    \widetilde{V}_{n,X}\widetilde{V}_{n,Y}^*\rangle\right) - {\rm
      Re}\left(\langle \widetilde{V}_{{\rm
        ant},X}\widetilde{V}_{n,Y}^*\rangle\right) - {\rm
      Re}\left(\langle \widetilde{V}_{{\rm
        ant},Y}\widetilde{V}_{n,X}^*\rangle\right) \right],\\
  V_{\rm cal}(t,\nu) ={} &i\Big[T_{{\rm ant},XY}(t,\nu) - T_{{\rm ant},YX}(t,\nu)\Big] \nonumber\\ ={}
  &\frac{-2}{k_B\Delta\nu}\left[\frac{{\rm Im}\left(\langle
      \widetilde{V}_X\widetilde{V}_Y^*\rangle\right)}{\sqrt{G_{T,X}G_{T,Y}}}
    - {\rm Im}\left(\langle
    \widetilde{V}_{n,X}\widetilde{V}_{n,Y}^*\rangle\right) - {\rm
      Im}\left(\langle \widetilde{V}_{{\rm
        ant},X}\widetilde{V}_{n,Y}^*\rangle\right) - {\rm
      Im}\left(\langle \widetilde{V}_{{\rm
        ant},Y}\widetilde{V}_{n,X}^*\rangle\right)\right],
 \label{eq:appdx_stokes_calUV}
 \end{flalign}
Since the sky signal ($\widetilde{V}_{\rm ant}$) and the electronic
noise ($\widetilde{V}_n$) are uncorrelated both within the same
polarization and among the two polarizations, time averaged values of
the cross terms cancel out. Hence, the calibrated Stokes parameters
reduce to Equations \eqref{eq:stokes_I_cal}-\eqref{eq:stokes_V_cal}.

\section{C. Supplementary CTP Measurements}
\label{appdx:lab_calibration}
\subsection{Calibration Data}
The CTP calibration equations are based on laboratory measurement of
the system's FE and receiver. Since the entire signal chain, from the
FE' input to the output at the coaxial cables before entering the ADC,
is considered as a single two-port network, two separate sets of $S$-
and noise-parameters were determined for both
polarizations. Additional details of the measurements and related
information can be found in \cite{nhan2018cosmic}.

\begin{figure}
  \centering
  \includegraphics[width=0.4\columnwidth]{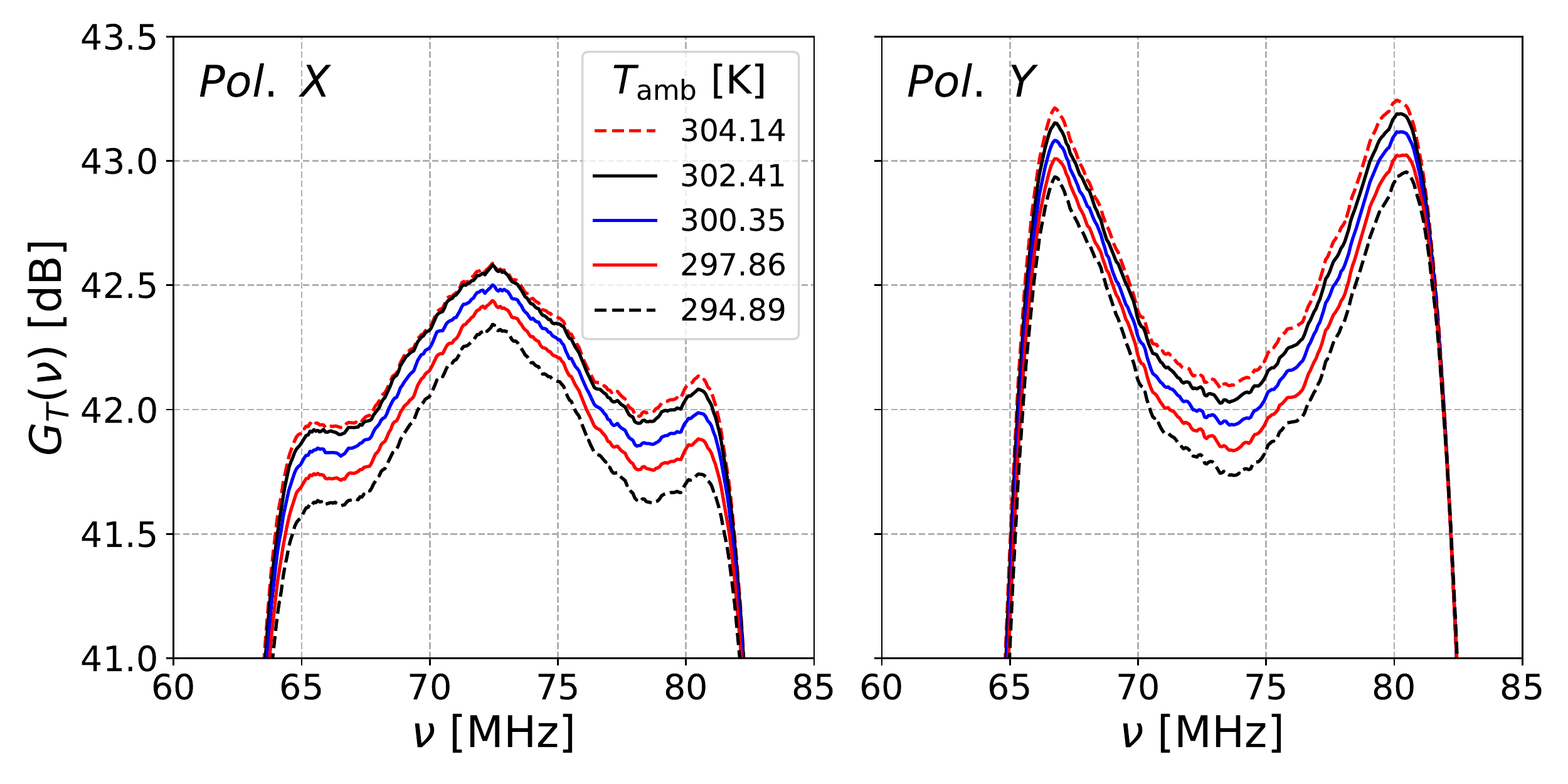}
  \quad\quad
  \includegraphics[width=0.3\columnwidth]{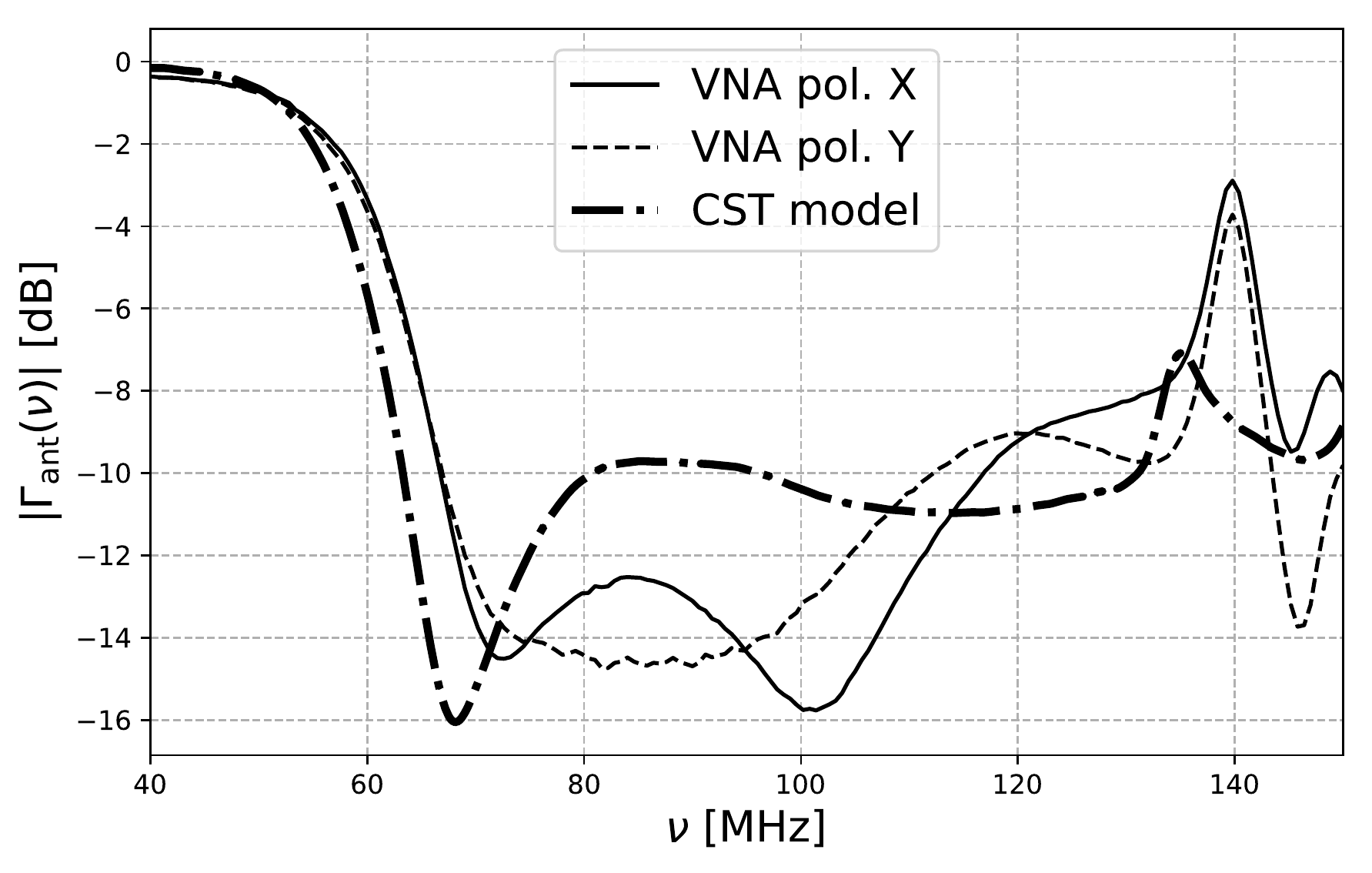}
  \caption{(Left) Laboratory measurement of the transducer gain of both
    polarizations as the operating temperature of the thermal
    enclosure is being varies. (Right) Antenna reflection coefficients
    measured as $S_{11}$ for both polarized and compared to the CST
    model. The difference between the measurement and model is due to
    the frequency response of the passive balun which was not included
    in the CST model. Nonetheless, the actual calibration utilized the
    direct $S_{11}$ measurement for each respective polarization, not
    the CST model.}
  \label{fig:trans_gain_correction}
\end{figure}

The $S$-parameters were measured with a VNA connector to the input and
output of the CTP's system at different set operating temperature
($T_{\rm amb} \sim 20-35^{\circ}$C). Then they are least-squares
fitted for a set of coefficients as functions of frequency and $T_{\rm
  amb}$. The corresponding transducer gain $G_T(\nu,T_{\rm amb})$ were
calculated with Equation \eqref{eq:transducer_gain_def}, using
$\Gamma_{\rm src} = 50\ \Omega$ in the laboratory
(\autoref{fig:trans_gain_correction}, left), and $\Gamma_{\rm src} =
\Gamma_{\rm ant}(\nu)$ (\autoref{fig:trans_gain_correction}, right)
for the CTP sleeved dipole in the field).

\begin{figure}
  \centering
  \includegraphics[width=0.315\columnwidth]{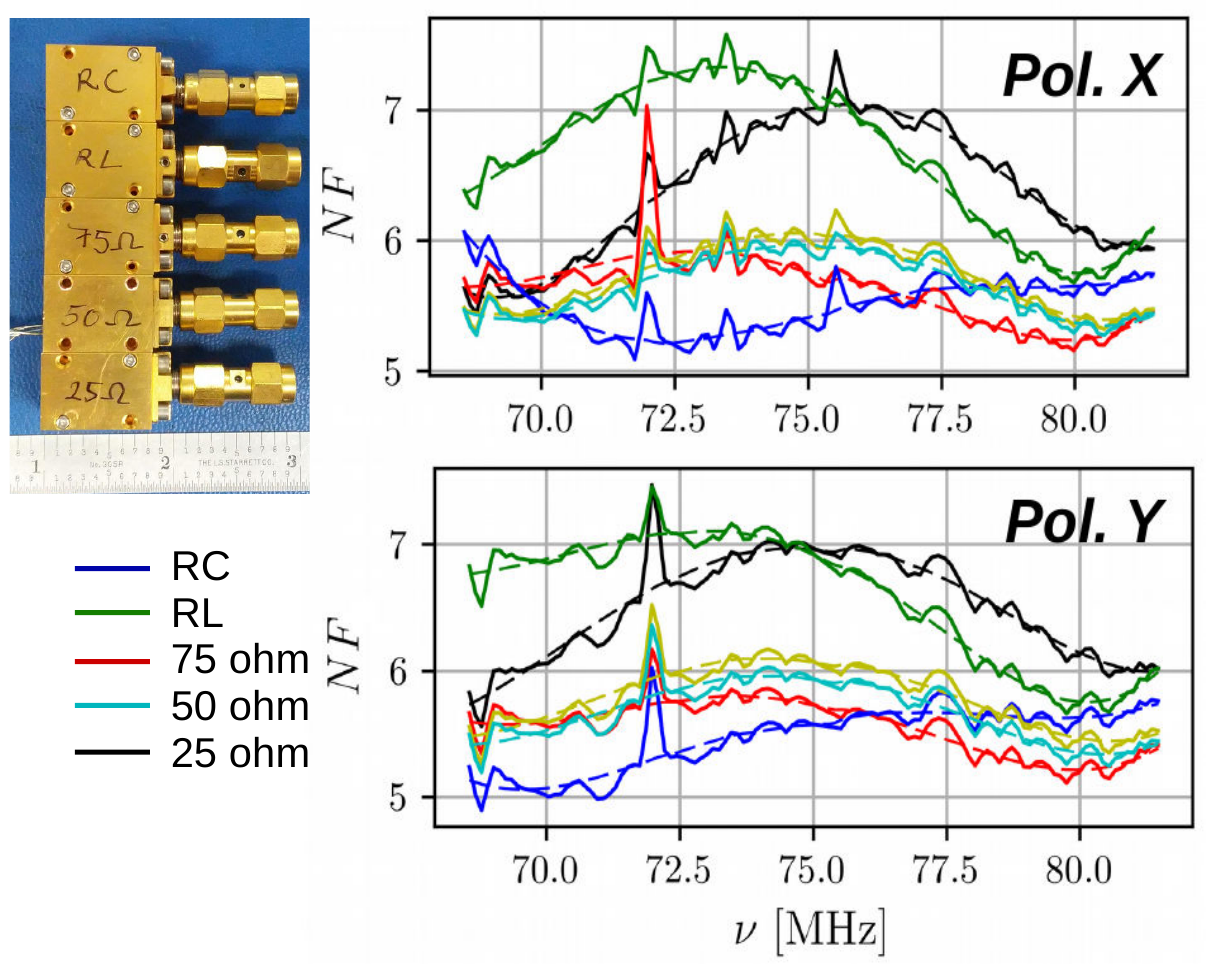}
  \quad\quad
  \includegraphics[width=0.45\columnwidth]{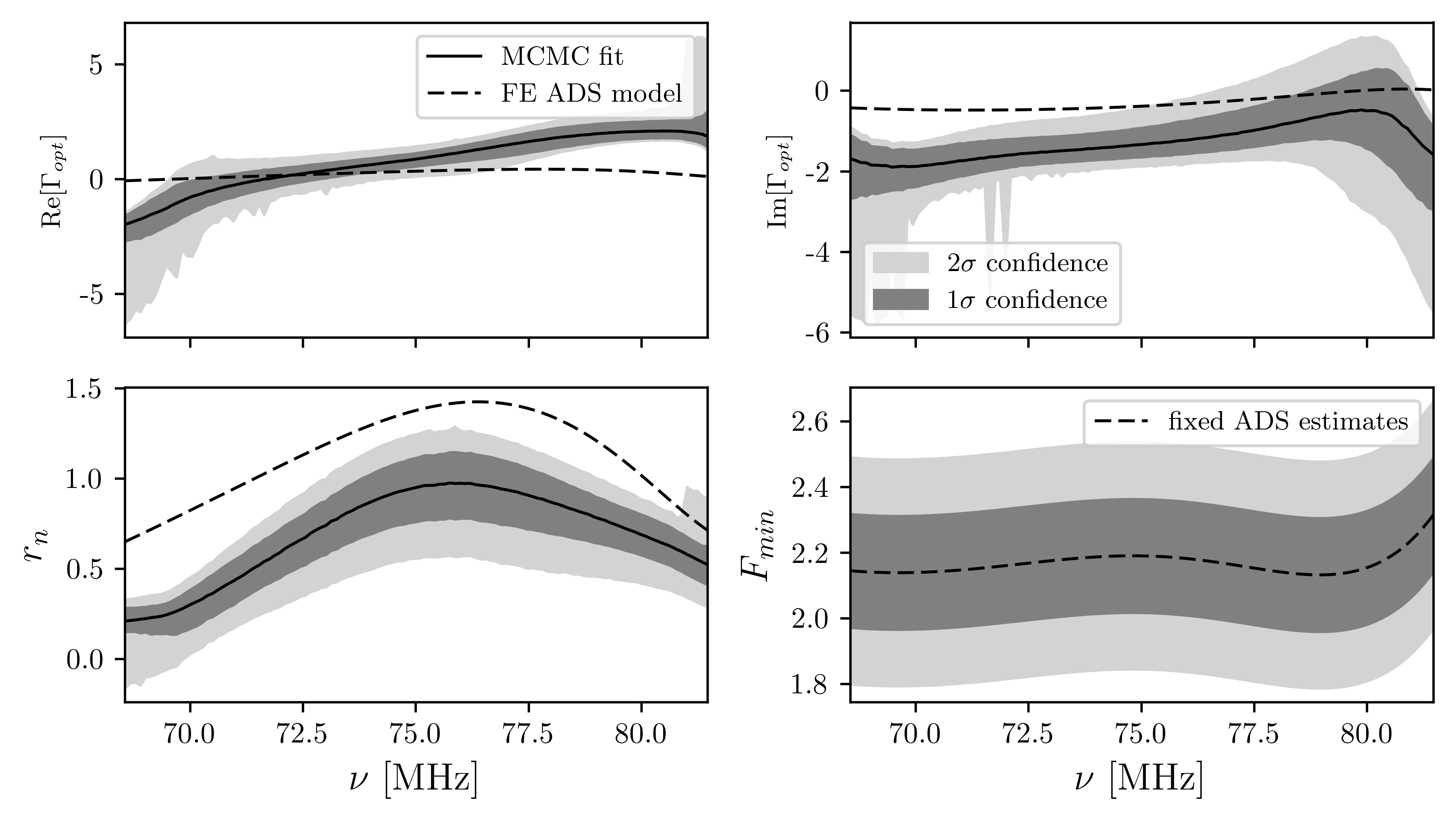}
  \caption{(a) Noise figures for polarization $X$ (upper panel) and
    $Y$ (lower panel) of the CTP system (except the sleeved antenna)
    measured for different input impedance modules, including
    $25\Omega,\ 50\Omega,\ 75\Omega,\ RC$, and $RL$ loads (solid). The
    best fit values of these $NF$ are superimposed (dashed). (b) MCMC
    fit of the noise parameters based on the lab measured $F_n(\nu)$
    and $\Gamma_{\rm src}(\nu)$, along with the ADS estimation of
    $F_{\rm min}(\nu)$ for the FE RF module. Since $F_{\rm min}(\nu$
    is a free constant, the MCMC fit does not converge unless it is
    constrained.}
  \label{fig:noise_temperature_correction}
\end{figure}

The CTP's noise parameters were determined by simultanesouly fit the
measured $T_n(\nu, Z_{\rm src})$ of the five reference input fixtures
to Equation \eqref{eq:noise_parameter}
(\autoref{fig:noise_temperature_correction}, left), where the noise
factor $NF(\nu) = 10\log_{10}F_n(\nu) = 10\log_{10}[1 +
  T_n(\nu)/T_0]$. However, since $T_{\rm min}(\nu)$ in that equation
is a free constant, the MCMC fit does not converge unless an initial
estimate of the $T_{\rm min}(\nu)$ is provided. For the CTP frontend,
by assuming $T_n(\nu)$ is dominated by the first statge LNA according
to Friis's formula, $T_{\rm min}(\nu)$ was constrained by a detailed
circuit model of the LNA using the ADS (Keysight's Advanced Design
Simulation) program. Subsequently, the remaining three noise
parameters were determined by the MCMC fit. The MCMC fit of
$\Big\{{\rm Re}[\Gamma_{\rm opt}(\nu)],\ {\rm Im}[\Gamma_{\rm
    opt}(\nu)],\ N(\nu)\Big\}$ are presented in the right panel of
\autoref{fig:noise_temperature_correction}, where $F_{\rm min}(\nu) =
1 + T_{\rm min}(\nu)/T_0$ is the noise factor, and $r_n(\nu) =
R_n(\nu)/Z_0$ is the normalized equivalent noise resistance with the
relation between $N(\nu)$ and $R_n(\nu)$ defined in
\eqref{eq:noise_parameter}.

\subsection{Effects of Savitzky–Golay Filter's Window Size}
\label{sec:savgol_win_sz_effect}
\begin{figure}[!htb]
\centering 
\includegraphics[width=0.575\columnwidth]{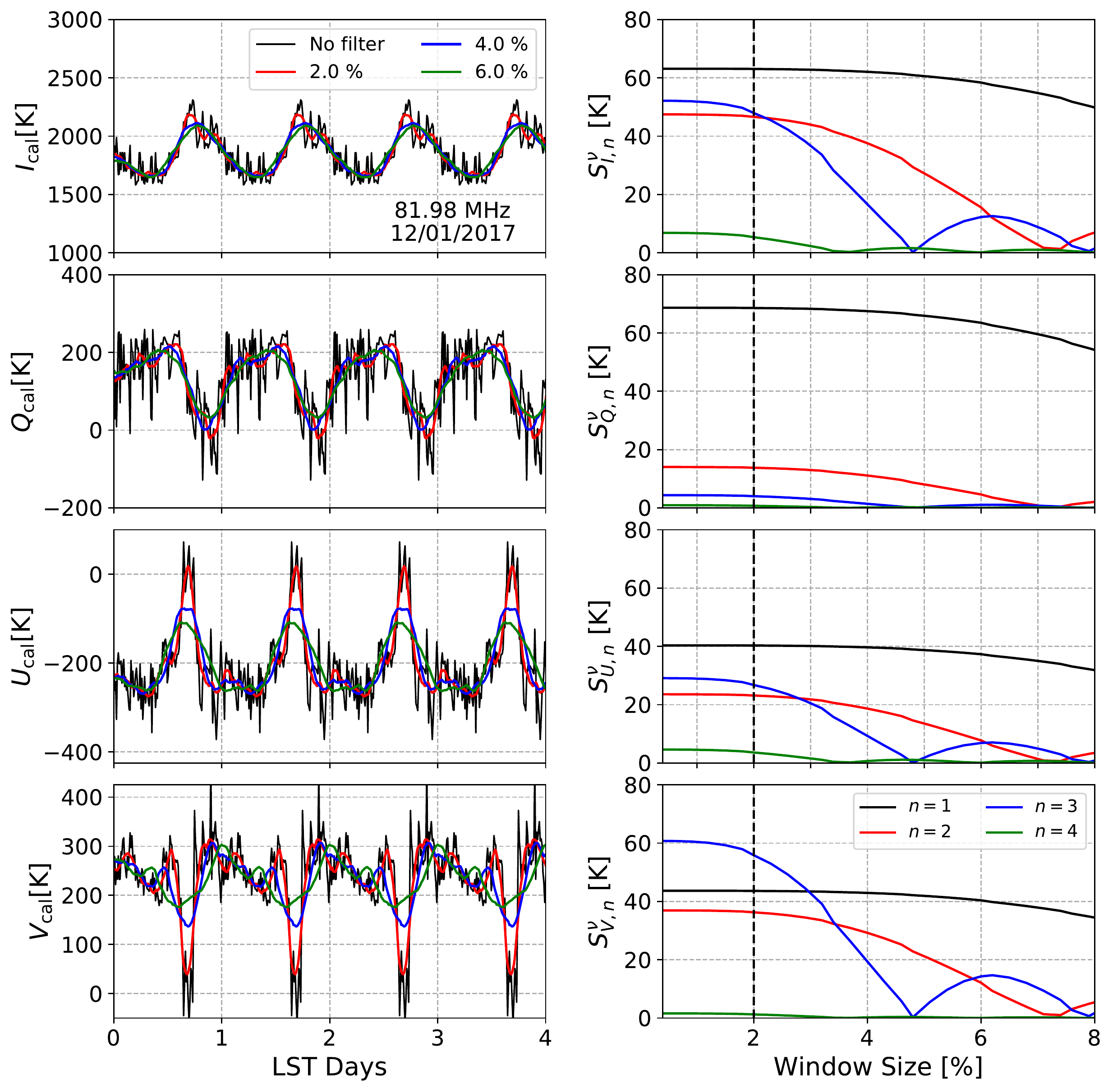}
\caption{Effects of Savitzky–Golay filter window size on the recovered
  magnitude of the harmonic components in the induced Stokes
  parameters. The filter computes the moving average of the Stokes
  parameters to suppress the high frequency noise. As a result, the
  larger the window size is, the more high-order harmonics are
  filtered out. (Left) The Stokes parameters of different window
  sizes: 2\%, 4\%, 6\% of the total data length compared with no
  filtering. (Right) Magnitudes of $n = 1,\ 2,\ 3,\ 4$ components
  decrease once the window size gets too large. The window size (in
  percentage relative to the data length) is chosen to be around 2\%
  of the total data length so that low-order harmonics of interest are
  preserved (dashed
  line).}\label{fig:ctp_obs_smth_win_sz_compare_82MHz_1nday_doy335}
\end{figure}

\nocite{*} \bibliographystyle{aasjournal} \bibliography{\myreferences}
\end{document}